# Self-similarity and scaling behavior of scale-free gravitational clustering

S. Colombi[1,2,4], F.R. Bouchet[2,4], and L. Hernquist[3,4,5]

[1] NASA/Fermilab Astrophysics Center, Fermi National Accelerator Laboratory,
P.O. Box 500, Batavia, IL 60510

[2] Institut d'Astrophysique de Paris, CNRS, 98 bis boulevard Arago, F-75014 - Paris, France.

[3] Board of Studies in Astronomy and Astrophysics, University of California,
Santa Cruz, CA 95064

[4] Institute for Theoretical Physics, University of California, Santa Barbara, CA 93106-4030

[5] Sloan Foundation Fellow, Presidential Faculty Fellow

**Abstract**

We measure the scaling properties of the probability distribution of the smoothed density field in $N$-body simulations of expanding universes with scale-free initial power-spectra, $\langle |\delta_k|^2 \rangle \propto k^n$, with particular attention to the predictions of the stable clustering hypothesis. We concentrate our analysis on the ratios $S_Q(\ell) \equiv \overline{\xi}_Q/\overline{\xi}_2^{Q-1}$, where $\overline{\xi}_Q$ is the averaged $Q$-body correlation function over a cell of radius $\ell$. According to the stable clustering hypothesis, $S_Q$ should not depend on scale. We measure directly the functions $S_Q(\ell)$ for $Q \leq 5$. The behavior of the higher order correlations is studied through that of the void probability distribution, $P_0$, which is the probability of finding an empty cell of radius $\ell$. If the stable clustering hypothesis applies, the function $P_0$ should also exhibit remarkable scaling properties.

In our analysis, we carefully account for various spurious effects, such as initial grid contamination, loss of dynamics due to the short range softening of the forces, and finite volume size of our simulations. Only after correcting for the latter do we find agreement of the measured $S_Q$, $3 \leq Q \leq 5$ with the expected self-similar solution $S_Q(\ell,t) = S_Q(\overline{\xi}_2) = S_Q(\ell/\ell_0(t))$, $\ell_0(t) \propto t^{4/(9+3n)}$. The void probability is only weakly sensitive to such defects and closely follows the expected self-similar behavior.

As functions of $\overline{\xi}_2$, the quantities $S_Q$, $3 \leq Q \leq 5$, exhibit two plateaus separated by a smooth transition around $\overline{\xi}_2 \sim 1$. In the weakly nonlinear regime, $\overline{\xi}_2 \lesssim 1$, the results are in reasonable agreement with the predictions of perturbation theory. In the nonlinear regime, $\overline{\xi}_2 > 1$, the function $S_Q(\overline{\xi}_2)$ is larger than in the weakly nonlinear regime, and increasingly so with $-n$. It is well-fitted by the expression $S_Q = (\overline{\xi}_2/100)^{0.045(Q-2)} \widetilde{S}_Q$ for all $n$. This weak dependence on scale proves *a small, but significant departure from the stable clustering predictions* at least for $n = 0$ and $n = +1$. It is thus also at variance with the predictions of the hierarchical model.

The analysis of $P_0$ confirms that the expected scale-invariance of the functions $S_Q$ is not exactly attained in the part of the nonlinear regime we probe, except possibly for $n = -2$ and marginally for $n = -1$. In these two cases, our measurements are not accurate enough to be discriminant. On the other hand, we could demonstrate that the observed power-law behavior of $S_Q$ cannot be generalized as such to arbitrary order in $Q$. Indeed this would induce scaling properties of $P_0$ incompatible with those measured.

**subject headings**: galaxies: clustering – methods: numerical – methods: statistical

# 1 Introduction

Large-scale structures in the observed galaxy distribution are thought to have arisen from small initial fluctuations through gravitational instability. Generally, it is assumed that the large-scale dynamics is dominated by collisionless dark matter. Of course, galaxy formation is not a collisionless process, and the extent to which one can infer some properties of the distribution of luminous matter from that of dark matter is a matter of debate. Moreover, statistical indicators in observational catalogs are subject to many contamination effects, such as the finiteness of the sampled volume, selection effects, redshift distortion in three-dimensional samples, and projection effects along the line of sight in two-dimensional sky surveys.

Albeit simpler than galaxy formation, the dynamics of collisionless matter on large scales in the Universe is still not fully-understood. The statistical evolution of self-gravitating collisionless particles may be described by the BBGKY hierarchy (e.g., Peebles 1980, hereafter LSS), which is an infinite system of coupled differential equations for the $Q$-body correlations functions in phase-space. In the weakly nonlinear regime, where the fluctuations of the density field are small, the hierarchy can be closed through perturbative approaches. However, it does not appear possible to solve the BBGKY hierarchy analytically when fluctuations of the density field become large, owing to the long-range nature of gravitational forces. In the highly nonlinear regime, therefore, $N$-body simulations are required to study the evolution of the density distribution and to measure the statistical properties of the system under consideration.

In this paper we analyze flat universes consisting entirely of collisionless matter and having scale-free initial Gaussian fluctuations

$$P(k) \equiv \langle |\delta_k|^2 \rangle = A \, k^n. \tag{1}$$

To analyze the statistics of gravitational clustering, we employ the count probability distribution function $P_N(m, \ell)$ (CPDF), which represents the probability of finding $N$ objects in a cell of radius $\ell$ located randomly in a sample of points with average number density $m$. The generating function $\mathcal{P}(\lambda) = \sum \lambda^N P_N$ of the CPDF can be written (White 1979; Schaeffer 1984; Balian & Schaeffer 1989a, hereafter BS; Szapudi & Szalay 1993)

$$\mathcal{P}(\lambda) = \exp\left\{ \sum_{Q=1}^{\infty} \frac{\left(\frac{4\pi}{3} n\ell^3\right)^Q (\lambda - 1)^Q}{Q!} \overline{\xi}_Q(\ell) \right\}, \tag{2}$$

where $\overline{\xi}_Q$ is the average of the $Q$-body correlation function $\xi_Q$ (e.g., LSS) over a cell:

$$\overline{\xi}_Q(\ell) \equiv \left[\frac{4\pi}{3}\ell\right]^{-3Q} \int_v d^3r_1 ... d^3r_Q \xi_Q(\mathbf{r}_1, ..., \mathbf{r}_Q). \tag{3}$$

The CPDF and its moments are thus well-suited to describing the scaling properties of a system of particles. From the dynamical point of view, an especially interesting quantity is the ratio

$$S_Q(\ell) \equiv \frac{\overline{\xi}_Q}{\overline{\xi}_2^{Q-1}}. \tag{4}$$

Indeed, as discussed below, $S_Q$ is expected to vary only weakly with scale or even to obey the following *scaling relation* over some range

$$S_Q = \text{constant of scale}, \tag{5}$$

as suggested by measurements of the low-order correlation functions $\xi_Q$ in the observed galaxy distribution (e.g., Groth & Peebles 1977; Fry & Peebles 1978; Davis & Peebles 1983; Sharp, Bonometto & Lucchin 1984; Szapudi, Szalay & Boschán 1992, Bouchet et al. 1993). Note that $S_3$ and $S_4$ are merely the (renormalized) skewness and kurtosis of the smoothed density distribution function.

In principle, it is straightforward to measure $S_Q$ because this quantity is simply related to the moments of the CPDF (e.g., Szapudi & Szalay 1993). However, for larger and larger $Q$, $S_Q$ is increasingly dominated by the high $N$ tail of the CPDF which is subject to spurious sampling effects related to the finite size of the volume (Colombi, Bouchet & Schaeffer 1994, hereafter CBSI). It is possible to account for these difficulties and accurately determine $S_Q$, but only for moderate orders $Q \lesssim 10$ (CBSI). In what follows, we limit our analysis to the fifth order, $Q = 5$. Higher-order statistics are considered indirectly through the void probability distribution function $P_0(m, \ell)$ (VPDF). From equation (2), it is indeed easy to write $P_0(m, \ell)$ as

$$P_0(m, \ell) = \exp\left[-\overline{N}\hat{\sigma}(m, \ell)\right], \tag{6}$$

where

$$\hat{\sigma}(n, \ell) = \sum_{N \geq 1}(-1)^{N-1}\frac{S_N(\ell)}{N!}N_c^{N-1}. \tag{7}$$

The quantity $\overline{N} \equiv (4/3)\pi m \ell^3$ is the average number of objects per cell, and $N_c$ is the typical number of objects in a cell located in an overdense region

$$N_c \equiv \overline{N}\,\overline{\xi}_2. \tag{8}$$

It is natural to study the scaling behavior of the VPDF by determining $\hat{\sigma}$ as a function of $N_c$. Indeed, if the scaling relation (5) applies, then (White 1979, BS)

$$\hat{\sigma}(n, \ell) = \sigma(N_c). \tag{9}$$

Measurements of the VPDF in the observed galaxy distribution are in good agreement with equation (9) (e.g., Sharp 1981; Bouchet & Lachièze-Rey 1986; Maurogordato & Lachièze-Rey 1986; Fry et al. 1989; Maurogordato, Schaeffer & da Costa 1992).

In the weakly nonlinear regime (WR), $\overline{\xi}_2 \ll 1$, the equations of motion can be solved perturbatively. For example, second-order perturbation theory (LSS) yields $\xi_3$ (Fry 1984a) and hence the skewness $S_3(\ell)$ of the smoothed density field (Juszkiewicz & Bouchet 1992; Juszkiewicz, Bouchet & Colombi 1993, hereafter JBC). Higher-order perturbation theory can be used to compute additional correlations (Goroff et al. 1986; Bernardeau 1992), such as the kurtosis $S_4(\ell)$ (Bernardeau 1994a; Lokas et al. 1994). In fact, the full hierarchy of averaged correlations $\overline{\xi}_Q(\ell)$ can be determined in the limit $\overline{\xi}_2 \ll 1$ (Bernardeau 1994b, hereafter B94). Note that the hierarchy of the ratios $S_Q$ depends only weakly on the cosmological parameters $\Omega$ and $\Lambda$, at least for $Q = 3$ (Bouchet et al. 1992, 1995; Hivon et al. 1995). As shown below, for scale-free initial conditions and $\Omega = 1$, the function $\overline{\xi}_2$ is a power-law of scale and $S_Q$ obeys the scaling relation (e.g., JBC, B94).

In the strongly nonlinear regime (SR), $\overline{\xi}_2 \gg 1$, no general analytical solutions to the BBGKY hierarchy have been found. Indeed, it is hard to deal with an a priori infinite hierarchy of correlations, since in this regime $\overline{\xi}_2 \ll \overline{\xi}_3 \ll ... \ll \overline{\xi}_Q$, so standard truncation methods cannot be applied. However, there is no preferred scale in gravitational dynamics and the BBGKY hierarchy admits self-similar solutions (e.g., Davis & Peebles 1977). Moreover, local statistical equilibrium should obtain on small enough scales, corresponding to virialized objects. In that case, the BBGKY hierarchy simplifies and the ratios $S_Q$ obey the scaling relation (e.g., Davis & Peebles 1977; Balian & Schaeffer 1989b). However, except in some particular frameworks (Fry 1984b, Hamilton 1988), the values of $S_Q$ in this regime are currently unknown and only measurements in $N$-body simulations can help to determine them. In particular, there is no reason why the values of $S_Q$ should be the same in the SR and in the WR.

In existing three-dimensional galaxy catalogs, the difference between these two regimes is weak, i.e. the property (5) seems to apply over all the available dynamic range (e.g., Bouchet et al. 1993), but this is likely to be due, at least partly, to projection in redshift space which tends to flatten the functions $S_Q(\ell)$ (Lahav et al. 1993; Matsubara & Suto 1994; Hivon et al. 1995). A bias between the distribution of galaxies and the mass could also flatten the $S_Q$. In two-dimensional galaxy catalogs, the $S_Q$ show a significant scale dependence (Gaztañaga 1994), but the interpretation of this result is complicated by projections effects.

Measurements of low-order correlations in $N$-body simulations indicate that $S_Q$ changes from the WR to the SR (e.g., Bouchet & Hernquist 1992, hereafter BH; CBSI; Lucchin et al. 1994). However, while the WR agreement between $N$-body simulations and perturbation theory is well-established (e.g., JBC; B94; Gaztañaga & Baugh 1995), the nonlinear regime behavior is still quite uncertain. The early $N$-body experiments of Efstathiou et al. (1988, hereafter EFWD) with scale-free initial conditions indicate that the $Q$-body correlations ($Q \leq 3$) exhibit self-similar and scale-invariant behavior in the SR, in agreement with Fry, Melott & Shandarin (1993). In a similar spirit, if the scaling relation applies, the CPDF itself should display remarkable scaling properties, as predicted by BS. Although testing these predictions is quite a delicate matter, the scaling relation can be strongly discriminated against in this way (CBSII). Bouchet, Schaeffer & Davis (1991, BSD) and BH measured the CPDF in $N$-body simulations with Cold Dark Matter (CDM), Hot Dark Matter (HDM) and white noise initial conditions and found striking agreement with the predictions of BS. CBSI measured the functions $S_Q(\ell)$, $Q \leq 5$ in the same simulations and found good agreement with scale-invariance in the highly nonlinear regime, but only after correcting for finite volume effects. However, there is not yet a consensus in the scientific community as to whether or not the scaling relation is reached in the highly nonlinear regime. For example, Lahav et al. (1993) and Suto & Matsubara (1994) who measured respectively the low-order moments of the CPDF and the functions $\xi_Q$, $Q \leq 4$ in low density CDM and white noise simulations, find a significant disagreement with equation (5) in the SR.

The measurement of correlations is difficult because the samples we have access to are not perfect. First, the finite volumes of the samples can induce unphysical distortions of the $S_Q$ (CBSI). Other spurious effects, such as discreteness, or artificial correlations related to a particular way of setting up initial conditions (CBSII), as well as other numerical effects due to some intrinsic properties of the $N$-body codes used to make the simulations can also introduce systematic errors. One has to carefully control all these defects before accepting any far-reaching conclusions.

In this paper, we carefully measure the quantities $S_Q(\ell)$ for $Q \leq 5$ and the VPDF in $N$-body simulations with scale-free initial power-spectra given by equation (1) with $n = -2, -1, 0, +1$. We explicitly account for the possible contamination effects noted above, particularly those related to the finite volume of the simulations. In addition to exploring the questions of whether or not the scaling-relation is obeyed in the highly nonlinear regime, our simulations enable us to study how the scaling behavior of the system changes with initial conditions.

The paper is organized as follows. In § 2, we describe the $N$-body simulations and how we measured the CPDF in our samples. The self-similar solution expected in our scale-free systems is recalled in § 3, along with possible spurious effects which can induce artificial deviations from such a solution. In § 4 we measure the two-body correlation function, compare the results with theoretical predictions, and determine a scale range where the effects noted in § 3 should be negligible or can be accounted for. We then measure the ratios $S_Q$, $Q \leq 5$ and show *the necessity* of correcting for finite volume effects. We show that agreement with perturbation theory is good in the WR. In the SR, the expected scale invariance does not seem to be exactly achieved, except perhaps for $n = -1$ and $n = -2$. The deviations from equation (5) are however very weak. Section § 5 deals with the void probability distribution. We test the agreement of the measured VPDF with the self-similar behavior and discuss all the possible contamination effects introduced in § 3. We can then carefully measure $\hat{\sigma}$ as a function of $N_c$ in the nonlinear regime and see if our conclusions for the low-order correlations can be generalized to higher order. In § 6, we summarize the results and conclude the paper with a short discussion.

## 2 The $N$-body samples

We performed a set of five scale-invariant simulations of flat universes involving $N_{\rm par} = 64^3$ dark matter particles in a cubic box of width $L_{\rm box} \equiv 1$ with periodic boundaries using the cosmological treecode of Hernquist, Bouchet & Suto (1991). The tolerance angle was $\theta = 0.75$ and the softening parameter $\varepsilon$ of the short range component of the force was given by $\varepsilon = \lambda_{\rm par}/20$, where

$$\lambda_{\rm par} = L_{\rm box} N_{\rm par}^{-1/3} \simeq 0.016 \qquad (10)$$

is the mean interparticle distance. The initial power-spectra are given by equation (1) with $n = -2$, $-1$, $0$ and $+1$. Two simulations with different initial random phases were performed for the case $n = -1$. Following EFWD, we chose the normalization factor $A$ so that the amplitude of the power-spectrum matched the white-noise level at the Nyquist frequency of the particles, except for $n = -2$. In that case, $A$ was taken four times smaller. The timestep choice is discussed in Appendix A. Table 1 lists the values of the expansion factor for which we have analyzed the $N$-body data.

For each simulation and each expansion factor analyzed, we measured the CPDF for spherical cells of radius $\ell$ in the following scale range (in units of $L_{\rm box}$)

$$-2.8 \leq \log_{10} \ell \leq -1.0, \qquad (11)$$

with a logarithmic step, $\Delta \log_{10} \ell = 0.2$. The lower limit $\ell_{\rm low}$ is imposed because we want to sample scales larger than the spatial resolution of the simulations, which should be a few $\varepsilon$. We took $\ell_{\rm low} \simeq 2\varepsilon$, but we shall see later that this constraint is not sufficiently strong. The upper scale $\ell_{\rm max}$ was chosen so that the number of independent modes in Fourier space corresponding to $\ell_{\rm max}$ is large. This does not guarantee, however, that finite volume effects are negligible, as we shall see later.

To compute the CPDF, we sample our simulations with a regular pattern of cells, involving $N_{\rm nodes}(\ell)$ points. Table 2 gives the chosen $N_{\rm nodes}$ for all the snapshots except for $(n, a) = (-2, 2)$, as a function of scale and the corresponding minimal value of the CPDF $P_N^{\rm min}$ that can be measured. It is important to realize that, to have an accurate sampling, the mean inter-cell separation must be much smaller than the scale of interest, which is the case with the choice of $N_{\rm nodes}$ given in Table 2. The statistical errors arising from the finite number of cells used to sample the $N$-body catalogs are quite small, in fact negligible compared with the other errors, such as those related to finite volume effects. For the case $(n, a) = (-2, 2)$, we used a smaller number of cells than for the other cases, $N_{\rm nodes} = 512^3$ for $\log_{10}\ell \leq -1.4$ (so we probably slightly undersample for $\log_{10}\ell = -2.8$), $N_{\rm nodes} = 384^3$ for $\log_{10}\ell = -1.2$ and $N_{\rm nodes} = 128^3$ for $\log_{10}\ell = -1.0$.

Table 1: Important scales in the simulations

| $n^{\rm a}$ | $a^{\rm b}$ | $\ell_0^{\rm c}$ | $s^{\rm d}$ | $\ell_c^{\rm e}$ | $\log_{10}\ell_{\rm m}^{\rm f}$ | $\log_{10}\ell_{\rm M}^{\rm g}$ | FV[h] | symbol[i] |
|---|---|---|---|---|---|---|---|---|
|  | 2.0 | 0.021 | 0.021 | 0.0048 | -2.0[#] | -1.0 | N | square |
| -2 | 3.2 | 0.051 | 0.052 | 0.0021 | -2.4[#,@] | -1.4* | Y | triangle |
|  | 5.2 | 0.103 | 0.138 | n.a. | -2.4[@] | -1.0 | Y | star |
|  | 8.0 | 0.173 | 0.327 | n.a. | -2.4[@] | -1.2[π] | Y | diamond |
|  | 2.5 | 0.022 | 0.022 | 0.0029 | -2.2[#] | -1.0 | N | square |
| -1 | 6.4 | 0.057 | 0.054 | n.a. | -2.4[@] | -1.4* | Y | triangle |
|  | 16 | 0.131 | 0.136 | n.a. | -2.4[@] | -1.0 | Y | star |
|  | 4.0 | 0.019 | 0.019 | 0.0022 | -2.4[#,@] | -1.0 | N | square |
| 0 | 16 | 0.048 | 0.047 | n.a. | -2.4[@] | -1.4* | Y | triangle |
|  | 64 | 0.118 | 0.120 | n.a. | -2.4[@] | -1.0 | Y | star |
|  | 6.4 | 0.020 | 0.020 | 0.0016 | -2.4[#] | -1.0 | N | square |
| +1 | 41 | 0.048 | 0.049 | n.a. | -2.6[@] | -1.4 | N | triangle |
|  | 256 | 0.119 | 0.123 | n.a. | -2.6[@] | -1.0 | Y | star |

[a] spectral index.
[b] expansion factor (assuming that initially $a = 1$).
[c] the measured correlation length.
[d] the correlation length according to the self-similar solution (scaled to $\ell_0$ at $a = 2.0, 2.5, 4.0$ and 6.4 respectively for $n = -2, -1, 0$ and $+1$).
[e] the typical distance between two matter particles in an overdense region, when measurable.
[f] minimum reliable scale. The indices indicate the effects imposing $\ell \geq \ell_{\rm m}$, [#] for grid effects (we impose $\ell > 1.5\ell_c$) and [@] for finite force resolution effects.
[g] maximum available reliable scale, when measuring and correcting for finite volume effects for the quantities $S_Q$, $Q = 3, 4, 5$. The indices indicate when some constraint has to be obeyed, * means that $\ell \geq \ell_0$ was imposed for the correction to be valid, [π] means that larger scales were two noisy to be able to meaningfully correct the quantities $S_Q$ for finite volume effects.
[h] indicates if finite volume effects (FV) have been corrected for (Y) or not (N) in figure 4.
[i] symbols used in the figures.

## 3 Self-similarity and numerical limitations

Since there are no preferred scales in the initial power-spectra, there should be only one relevant physical scale in the simulated system: the correlation length $\ell_0$ defined by

$$\overline{\xi}_2(\ell_0) \equiv 1. \tag{12}$$

This scale separates the weakly nonlinear regime (WR) where $\overline{\xi}_2 \ll 1$ from the strongly nonlinear regime (SR) where $\overline{\xi}_2 \gg 1$. Any statistical quantity $f(\ell, t)$ can thus be written as

$$f(\ell, t) = g(\ell/\ell_0). \tag{13}$$

Table 2: Resolution of the sampling cells pattern as a function of logarithm of scale and the corresponding minimal CPDF reached

| $\log_{10} \ell$ | -2.8 | -2.6 | -2.4 | -2.2 | -2.0 | -1.8 | -1.6 | -1.4 | -1.2 | -1.0 |
|---|---|---|---|---|---|---|---|---|---|---|
| $N_{\text{nodes}}^{1/3}$ | 2176 | 1792 | 1536 | 1280 | 1024 | 896 | 768 | 640 | 512 | 384 |
| $\log_{10} P_N^{\min}$ | -10.0 | -9.75 | -9.55 | -9.3 | -9.0 | -8.85 | -8.65 | -8.4 | -8.1 | -7.75 |

In the WR, the two body correlation function is $\overline{\xi}_2 = C(n) a^2 \ell^{n-3}$, where $C$ is a constant depending only on the initial power spectrum index (see § 4.1.1 for the exact expression of $\overline{\xi}_2$), which implies that (Inagaki 1976, Davis & Peebles 1977, LSS)

$$\ell_0 = s(t), \quad s(t) \propto a^\alpha, \tag{14}$$

with

$$\alpha = \frac{2}{3+n}. \tag{15}$$

The third column of Table 1 gives the measured correlation length in units of $L_{\text{box}}$, to be compared to the values of $s(t)$. The latter are given in the fourth column of Table 1, assuming that the initial value of $s(t)$ is exactly equal to the measured $\ell_0$ at the smallest sampled expansion factor. The agreement of the measurements with equation (14) is very good, except for the case $n = -2$, where, at late stages, the deviation from self-similar behavior appears at first sight quite significant (for $a \gtrsim 5.2$). We shall see in § 4.1.3 that this lack of agreement with self-similarity is spurious and is due to finite volume effects.

Deviations from the expected behavior given by equations (13) and (14) should only be due to numerical limitations of $N$-body simulations, unless linear theory is not valid in the WR. Here, as in Hivon et al. (1994), we list numerical limitations of $N$-body simulations that can contaminate the measurements. In § 3.1, we discuss possible dynamical effects due to the discreteness of our representation of an underlying continuous density field. § 3.2 deals with grid effects and transients resulting from the use of the Zel'dovich approximation to slightly perturb a regular pattern of particles to set up initial conditions. In § 3.3 we mention a possible lack of accuracy at small scales due to the short range softening of the forces in the simulations. In § 3.4, we discuss finite volume effects.

## 3.1 Collisionless fluid limit

The $N$-body code should describe the evolution of a "self-gravitating" system of "micro"-particles. On the scales of interest, the collisionless (mean-field limit) Boltzman equation, i.e. the Vlasov equation, is valid (see, e.g., LSS). The "macro"-particles we use in $N$-body codes have masses much larger than those of the micro-particles. Even if the real underlying distribution is well approximated by a mean-field description, it may not be the case for our system of macro-particles in the scaling regime we consider. To reduce dynamical effects arising from the discreteness of our representation, such as two-body relaxation, a short range softening parameter $\varepsilon$ is introduced (see § 3.3, Appendix A). In itself, however, this is not sufficient. Before making measurements on the $N$-body simulation, one must wait long enough for the numerical system to indeed achieve the mean-field limit of a "collisionless fluid". In other words, the typical number of particles per collapsed object (of size similar to the Jeans length) must be large. The typical size of a clump is the correlation length. Thus, we should have

$$\ell_0 \gg \lambda_{\rm par} \simeq 0.016. \tag{16}$$

This condition is fulfilled for the expansion factors we chose to analyze, although barely for the smallest ones (corresponding to $a = 2.0$, 2.5, 4.0 and 6.4 for $n = -2, -1, 0$ and 1 respectively, see Table 1).

## 3.2 Initial conditions

To reduce the small-scale shot-noise arising from the discreteness of our "macro"-particles, one traditionally starts the simulations from a regular distribution of particles, slightly perturbed by using the Zel'dovich approximation (Zel'dovich 1970) to displace the particles. Some imprint of this initial pattern is conserved during the simulation, particularly in underdense regions. It can thus contaminate the measurement of the CPDF, especially at small $N$ (e.g., BSD, BH, CBSII). From the statistical point of view, the discrete realization of a continuous density field should indeed be locally Poissonian. We shall discuss in more detail such "grid" effects on the VPDF in § 5. As far as the low-order correlation functions are concerned, these effects should be negligible, except in the early stages of the simulation, at scales smaller than, or of the same order as the mean interparticle distance $\ell_c$ in overdense regions.

The low-order correlations are dominated by regions with high density contrasts, and increasingly so with $Q$. Because nonlinear gravitational dynamics is intrinsically chaotic, it is likely that such regions become locally Poissonian after shell-crossing. To have such phase-space mixing, condition (16) must be fulfilled, but it is not obvious that this is a sufficient condition: it is possible to have high density contrasts without shell-crossing, particularly if there is a cut-off at small scales in the spectrum of initial fluctuations. Note that since the ratio between small-scale power and large-scale power decreases with $-n$, we expect grid effects to be less important for large values of $n$. They should also become less and less significant as the system evolves.

When the typical number of particles per cell located in an overdense region is large compared to unity, grid effects should be negligible. This condition yields a constraint on the scales which are fully reliable of the form

$$\ell \gg \ell_{\rm c}. \tag{17}$$

The scale $\ell_c$ is the typical distance between two particles in a cluster, defined by (BS)

$$N_{\rm c}(\ell_{\rm c}) \equiv 1. \tag{18}$$

Some measured values of $\ell_c$, when available, are given in Table 1. We transformed the asymptotic constraint (17) in the more practical condition

$$\ell > 1.5 \; \ell_{\rm c}. \tag{19}$$

We see for example that, for the first snapshot, we should have $\log_{10} \ell > -2.6$, $-2.50$, $-2.35$ and $-2.15$ respectively for $n = 1, 0, -1, -2$ for grid effects to have a negligible influence on the low-order correlations. This partly explains our choice of the "minimum reliable scale" $\ell_{\rm m}$ given in Table 1. For larger expansion factors, the measured clustering number $N_c$ is always sufficiently large in the available dynamic range (11) for grid effects to be insignificant (except for $(n, a) = (-2, 3.2)$ where we should also have $\log_{10} \ell > -2.5$).

Figure 1: The CPDF measured in the most evolved snapshot of our simulations as a function of $N$, for various scales $\ell$. The system of coordinates has been chosen to emphasize the large-$N$ tail of the CPDF, which is clearly exponential, although rather noisy. As $N$ increases, the CPDF indeed presents larger and larger irregularities followed by a sharp cutoff. The smooth lines are the analytical fits we derive in § 4.2.2. In each panel, the top curve corresponds to the scale $\log_{10} \ell = -1.0$. Going downwards, the scale decreases with each curve, with a logarithmic step $\Delta \log_{10} \ell = 0.2$.

The use of the Zel'dovich approximation to set up initial conditions can also induce some spurious effects (e.g., JBC). Indeed, this approximation describes a first (linear) order approximation to the trajectories of the particles. It is not valid if one aims at measuring higher order quantities, such as the skewness $S_3$ of the density distribution. It is thus necessary to wait sufficiently long that higher-order coupling terms had time to fully develop. The corresponding requirement is $a \gg 1$, which is hardly fulfilled by our simulations. Practically, however, $a \gtrsim 3$ should be sufficient (e.g., Baugh, Gaztañaga, & Efstathiou, 1994). We thus expect the first snapshot of our simulations to be contaminated by transients, particularly on large, weakly nonlinear scales. Note that the function $\overline{\xi}_2$, which is of first order in the WR, should be only weakly influenced by these effects.

### 3.3 Short range softening of the forces

The dynamics on the smallest scales in our simulations is not accurate because of the softening parameter $\varepsilon$ used to bound forces and reduce two-body relaxation (§ 3.1). Typically, however, if the timestep is carefully chosen (see, e.g., Appendix A), the softening parameter should contaminate measurements up to at most a few $\varepsilon$. We shall see, however, that softening can affect the count-in-cells statistics on scales as large as $\log_{10} \ell \sim -2.5$ (§ 4.1.2), which partly explains our choice of the "minimal reliable scale" $\ell_m$ in Table 1.

### 3.4 Finite volume effects

Since we use periodic boundaries in our simulations, the fluctuations from scales larger than the box size are missing. Also, on scales smaller than but comparable to the box size, only a few independent modes of the power spectrum are sampled, thereby increasing the uncertainties in measurements as one approaches the box size. To minimize these effects, one usually requires $\log_{10} \ell \lesssim -1$, as we did. But this constraint is not necessarily sufficient. Indeed, the large $N$ tail of the CPDF is determined by just a few large clusters and is thus subject to fluctuations due to small number statistics, until it reaches an artificial cutoff at $N_{\max}$ (CBSI, CBSII), as illustrated by Figure 1. If the sample size $L_{\text{box}}$ is large enough compared to the correlation length (typically $L_{\text{box}} \gtrsim 20\ell_0$), which is the typical size of a cluster, these modifications of the true tail occur for such small values of the CPDF that they are of no consequence. When the correlation length becomes significant compared to $L_{\text{box}}$ (typically $\ell_0 \gtrsim L_{\text{box}}/20$), this scale-dependent effect is likely to influence measurements of the low-order correlation functions $\overline{\xi}_Q$, increasingly with $Q$. Also, this finite volume effect should be more significant when the amplitude of the fluctuations on large scales is big, so it should increase with $-n$. Fortunately, it can be corrected for, or at least an error-bar associated with it can be evaluated (CBSI).

# 4 Scaling behavior of the low-order correlations

## 4.1 The averaged two-body correlation function

### 4.1.1 Theoretical predictions

In the WR, the function $\overline{\xi}_2$ can be written (e.g., Hivon et al. 1994)

$$\overline{\xi}_2(\ell) = \frac{9 \cdot 2^{n-1}}{4\pi} \frac{\Gamma(1-n)\Gamma\left(\frac{3+n}{2}\right)}{\Gamma^2\left(\frac{2-n}{2}\right)\Gamma\left(\frac{5-n}{2}\right)} A a^2 \ell^{-n-3}, \quad \ell \gg \ell_0 \tag{20}$$

where $A$ is the initial amplitude of the power-spectrum and $a$ is the expansion factor, assumed to be unity at the beginning of the simulation.

In the SR, with the additional hypothesis of local statistical equilibrium, the two-body correlation function should be a power-law of scale (Davis & Peebles 1977):

$$\overline{\xi}_2(\ell) \propto \ell^{-\gamma}, \quad \ell \ll \ell_0, \tag{21}$$

with

$$\gamma = \frac{3(3+n)}{5+n}. \tag{22}$$

### 4.1.2 Measurement

Figure 2 displays the quantity $\overline{\xi}_2$ as a function of $\ell/s(t)$ (where $s(t)$ is given by eq. [14]) in logarithmic coordinates and for the various expansion factors listed in Table 1. To compute $\overline{\xi}_2$ we use the following formula (Fry & Peebles 1978, LSS), that corrects for discreteness

$$\overline{N}^2 \overline{\xi}_2(\ell) = \mu_2 - \overline{N}, \tag{23}$$

where $\mu_2$ is the centered, second moment of the CPDF. More generally, the centered moment of order $Q$ is defined by

$$\mu_Q(\ell) \equiv \langle (N - \overline{N})^Q \rangle = \sum_{N=0}^{\infty} (N - \overline{N})^Q P_N(\ell). \tag{24}$$

We show results for only one of the two $n = -1$ simulations. Indeed, the measured $\overline{\xi}_2$ are nearly identical in the two simulations, except perhaps for the largest expansion factor $a = 16$, for which the differences are anyway smaller than $\Delta \overline{\xi}_2 / \overline{\xi}_2 \lesssim 0.3$. This number corresponds to a vertical length in figure 2 approximately 1.5 times smaller than the size of the symbols used to make the plots.

As displayed in Figure 2, if the evolution of the system is self-similar, all the curves should superimpose. Possible deviations from such behavior must be induced by numerical effects. The short dashed lines indicate the logarithmic slope expected from equations (21) and (22) in the SR. The normalization has been chosen to match the measurements. The long dashes give the linear theory prediction (20), expected to apply in the WR.

Except for the case $n = -2$, which we discuss further, the measured $\overline{\xi}_2(\ell, t)$ is in very good agreement with theoretical predictions. Deviations from self-similarity are only significant on the smallest scales, typically for $\log_{10} \ell \lesssim \ell_m$, with

$$\ell_m \simeq -2.5, \tag{25}$$

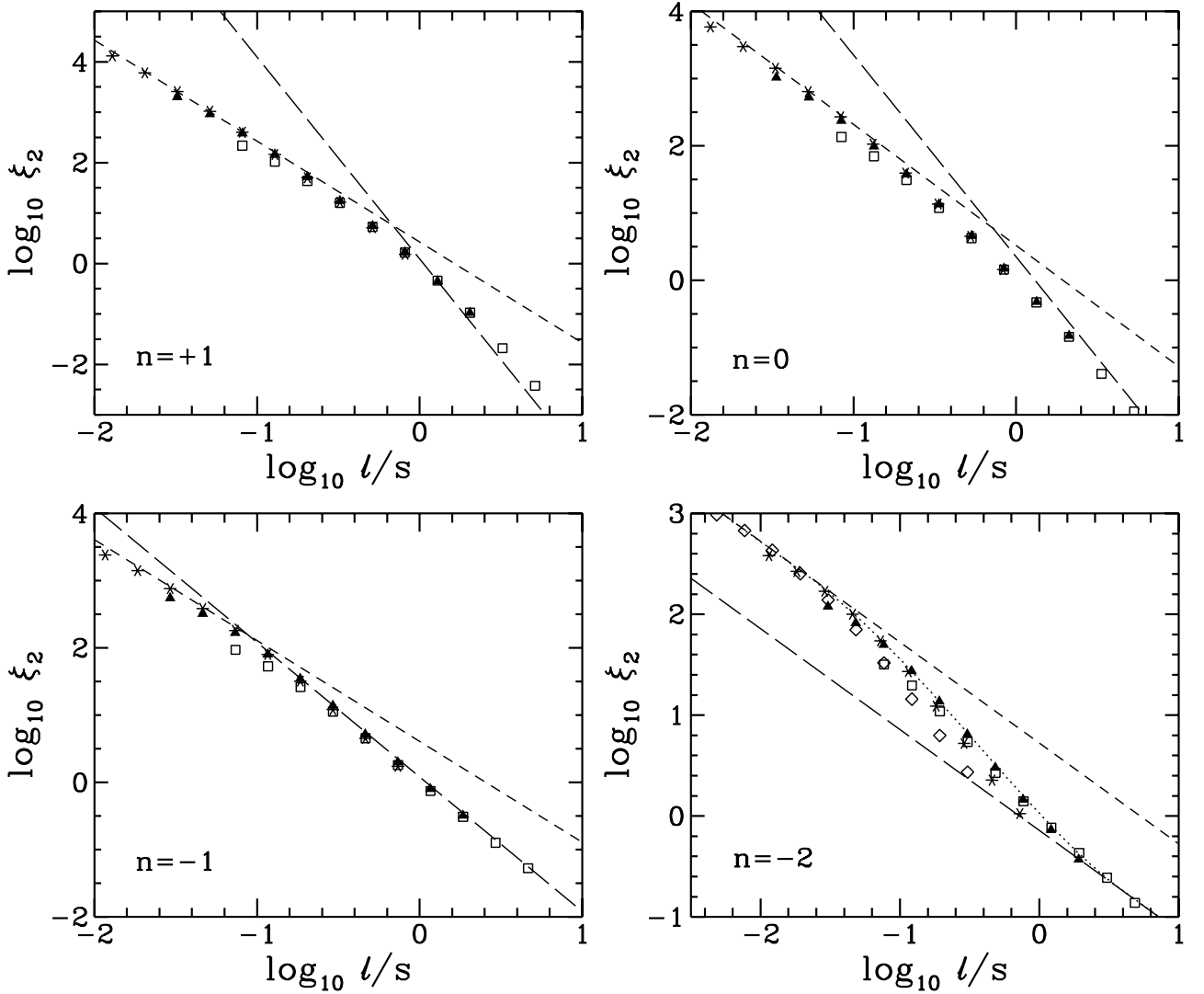

Figure 2: The averaged two-body correlation function $\bar{\xi}_2$ as a function of $\ell/s$ (where $s$ is given by eq. [14]) in logarithmic coordinates, measured in each simulation for various expansion factors (Table 1; see also this table for the significance of the symbols). The long dashes give the linear prediction, valid in the limit $\bar{\xi}_2 \ll 1$, and the short dashes indicate the logarithmic slope of $\bar{\xi}_2$ expected in the limit $\bar{\xi}_2 \gg 1$ (eqs. [21] and [22]). In the last case, the normalization has been chosen to fit the data. The dotted curve in the bottom right panel is a polynomial fit in logarithmic coordinates (see eqs. [27] and [28]).

which corresponds to the two left-most points of each curve in each panel of Figure 2, for which the measured $\overline{\xi}_2$ underestimates the true value. This effect seems to increase with $-n$ and tends to decrease at late stages in the simulations. For the first expansion factor analyzed in each simulation (or the first two for $n = -2$), the deviation is mainly due to the numerical limitations discussed in § 3.1, 3.2 and 3.3, particularly grid effects. For larger expansion factors, grid effects become negligible and only the short range softening of the forces introduced in § 3.3 should matter.

A careful measurement of the two-body correlation function $\xi_2(r)$ shows that at late stages in the simulations the system exhibits self-similar behavior on scales as small as $\ell_\varepsilon \simeq 2\varepsilon$. This excellent agreement with theoretical predictions does not apply for velocity correlations, which are much more sensitive to the force resolution. Note that the function $\overline{\xi}_2(\ell)$ is related to $\xi_2(r)$ through an integral over a cell (eq. [3]). In this integral, small separations $|\mathbf{r}_1 - \mathbf{r}_2| \lesssim \ell_\varepsilon$ can contribute significantly. Only when $\ell$ is large enough compared to $\ell_\varepsilon$, will the effect of softening become negligible. Experimentally, we find in equation (25) that $\overline{\xi}_2(\ell)$ can be contaminated up to the scale $\ell_{\rm m} \simeq 2\ell_\varepsilon \simeq 4\varepsilon$. Table 1 gives more accurate estimates of the "minimum reliable scale" $\ell_{\rm m}$ we chose.

### 4.1.3  Finite volume effects

The very good agreement with a self-similar behavior on scales larger than $\ell_{\rm m}$, and the convergence to the predictions of linear theory on large scales[1] indirectly suggest that finite volume effects on $\overline{\xi}_2$ are negligible in the scaling range we consider, for $n = 1$, 0 and $-1$. This can also be checked in a different manner with the two-body correlation: as discussed in CBSI, because we miss some power on scales larger than the simulation box size, the measurements of $\overline{\xi}_2(\ell)$ are likely to underestimate its true value, increasingly with $\ell$. In a box of finite volume, the measured function $\xi_2(\ell)$ can be written

$$\overline{\xi}_2^{\rm box}(\ell) \sim \frac{L_{\rm box}^3}{2\pi^2} \int_{\left(\frac{6}{\pi}\right)^{1/3}\frac{2\pi}{L_{\rm box}}}^{\infty} \langle |\delta_k^{\rm box}|^2 \rangle \left[W_\ell(k)\right]^2 k^2 dk, \qquad (26)$$

where $\langle |\delta_k^{\rm box}|^2 \rangle$ is the power spectrum measured in the box and $W_\ell \equiv 3(k\ell)^{-3}[\sin(k\ell) - k\ell\cos(k\ell)]$ is the Fourier transform of the top hat filter. One can correct such measurements for finite volume effects by extrapolating the measured power-spectrum $\langle |\delta_k^{\rm box}|^2 \rangle$ on scales larger than $L_{\rm box}$ using linear perturbation theory and by computing the integral (26) in the limit $L_{\rm box} \to \infty$. Of course, such a correction is valid as long as the large-scale dynamics does not dramatically influence the power-spectrum on scales smaller than the box, so the correlation length should be a reasonably small fraction of the box size. We applied the above procedure to our data and indeed noticed that it yields only small changes for $n > -2$. The correction is significant for $n = -2$ for all $a$ when $\log_{10} \ell > -1.4$, but does not increase $\overline{\xi}_2$ by more than a factor of order of $10^{0.1}$ at $\log_{10} \ell = -1.0$. This result is in agreement with the bottom right panel of figure 2 for $a = 2$ and $a = 3.2$. But for $a = 5.2$ and $a = 8.0$, it appears that finite volume effects are much stronger than our correction seems to indicate. However, in that case, our prescription is unreliable, because the correlation length becomes a significant fraction of the box size, larger than $L_{\rm box}/10$, so nonlinear coupling can affect Fourier modes with wavelengths as large as the size of the box or even larger, particularly at $a = 8$.

The strong deviation from self-similarity on scales larger than $\ell_{\rm m}$ in the $n = -2$ case is thus due to finite volume effects. The true function $\overline{\xi}_2$ is larger than the measured one, increasingly with scale. The correction invoked above shows however that one can reasonably infer from Figure 2 the global shape of $\overline{\xi}_2$, assuming that at each scale the true value of $\overline{\xi}_2$ is approximated (within a factor of $10^{0.1}$) by the maximum of the values obtained in each of the snapshots analyzed. The corresponding dotted curve in Figure 2 is given by the following function $F(x)$, in logarithmic coordinates

$$\begin{aligned} F(x) &= a_1(x - x_1) + y_1, & x &\leq x_1, \\ F(x) &= P_3(x), & x_1 &< x \leq x_2, \\ F(x) &= a_2(x - x_2) + y_2, & x &\geq x_2, \end{aligned} \quad (27)$$

where $P_3(x)$ is the polynomial of degree four satisfying

$$P_3(x_1) = y_1, \quad dP_3/dx_1 = a_1, \quad P_3(x_2) = y_2, \quad dP_3/dx_2 = a_2, \quad P_3(x_3) = y_3, \quad (28)$$

with $x_1 = -1.7$, $y_1 = 2.42$, $a_1 = -1$, $x_2 = 0.6$, $y_2 = -7.4$, $a_2 = -1$, $x_3 = -0.7$, $y_3 = 1.11$.

---

[1] except for $n = +1$: in this case, nonlinear effects might never be negligible, whatever scales are considered.

## 4.2 Higher order correlation functions

### 4.2.1 Theoretical predictions

The measurement of $\overline{\xi}_2$ allows us to define a range of scales for which the $N$-body simulations for $n > -2$ exhibit the expected self-similar behavior given by equations (13) and (14). Furthermore, in the SR, $\overline{\xi}_2$ is compatible with the predictions obtained when local statistical equilibrium is assumed. If this latter hypothesis is valid, the quantities $S_Q$ should obey the scaling property (5) in the regime $\overline{\xi}_2 \gg 1$. In the WR regime, the quantities $S_Q$ are expected to match the predictions of perturbation theory, i.e.,

$$S_3 = \frac{34}{7} - (n+3), \quad (29)$$

(Juszkiewicz & Bouchet 1992, JBC), and (B94)

$$S_4 = \frac{60712}{1323} - \frac{62}{3}(n+3) + \frac{7}{3}(n+3)^2, \quad (30)$$

$$S_5 = \frac{200575880}{305613} - \frac{1847200}{3969}(n+3) + \frac{6940}{63}(n+3)^2 - \frac{235}{27}(n+3)^3. \quad (31)$$

### 4.2.2 Measurements: correcting for finite volume effects

Figure 3 displays the ratios $S_Q \equiv \overline{\xi}_Q / \overline{\xi}_2^{Q-1}$ as functions of $\overline{\xi}_2$ as measured in our simulations, in the available scale range given by equation (11). To compute the averaged correlation functions $\overline{\xi}_Q$, we use the standard formulae, that correct for discreteness (Fry & Peebles 1978, LSS, CBSI) and generalize to higher orders from equation (23):

$$\overline{N}^3 \overline{\xi}_3(\ell) = \mu_3 - 3\mu_2 + 2\overline{N}, \quad (32)$$

$$\overline{N}^4 \overline{\xi}_4(\ell) = \mu_4 - 6\mu_3 - 3\mu_2^2 + 11\mu_2 - 6\overline{N}, \quad (33)$$

$$\overline{N}^5 \overline{\xi}_5(\ell) = \mu_5 - 10\mu_4 - 10\mu_2\mu_3 + 35\mu_3 + 30\mu_2^2 - 50\mu_2 + 24\overline{N}. \quad (34)$$

The quantity $\mu_Q$ is the centered moment of order $Q$ of the CPDF (eq. [24]).

Figure 3 shows that the $S_Q$ do not obey self-similarity as well as the averaged two-body correlation function, and it becomes worse with decreasing $n$. All the curves corresponding to various expansion factors $a$ but the same value of $Q$ (at fixed $n$) should superimpose, which is far from being the case, particularly for $n = -2$. At a fixed value of $\overline{\xi}_2$, the measured value of $S_Q$ decreases with time. These effects are spurious, due to the finiteness of the sampled volume. As discussed by CBSI and in § 3.4, finite volume effects are more important when the correlation length is a large fraction of the sample size and when there is more power at large scales. They induce increasing irregularities with $N$ on the CPDF until a sharp cutoff at $N = N_{\max}(\ell)$ is encountered, as illustrated by Figure 1. These irregularities, and particularly the cutoff at $N_{\max}$ contaminate the measurement of the low-order averaged correlation functions, which are increasingly sensitive to the large-$N$ part of the CPDF as $Q$ increases.

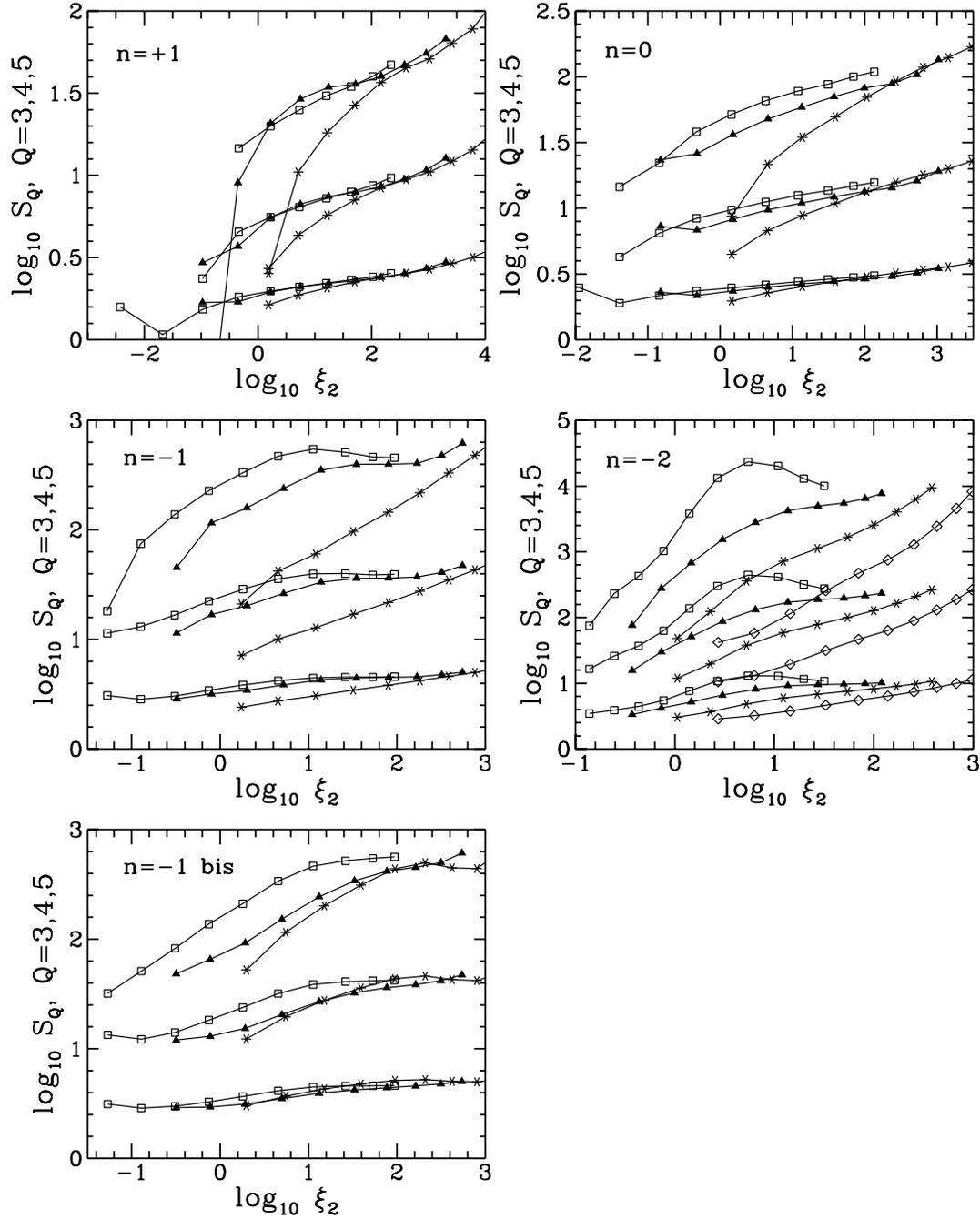

Figure 3: The quantities $S_Q \equiv \overline{\xi}_Q/\overline{\xi}_2^{Q-1}$ as functions of $\overline{\xi}_2$, in logarithmic coordinates, measured in each simulation for various expansion factors (Table 1; see also this table for the meaning of the symbols) in the full available scaling range (eq. [11]). In each panel, the values of $S_Q$ increase with the order $Q$. In this coordinate system, since there is no preferred scale, all the curves (at fixed $Q$ and $n$) should superimpose, which is far from being the case, even more so as $-n$ increases.

Fortunately, it is possible to use the procedure proposed by CBSI to correct for finite volume effects, by smoothing and extending to infinity the large-$N$ exponential tail exhibited by the CPDF. We fit such a tail with the following form (see Fig. 1)

$$P_N(\ell) \sim A(\ell) U(N/N_c) N^{\eta(\ell)} e^{-\beta(\ell) N}, \tag{35}$$

where

$$\beta(\ell) = |y_s(\ell)|/N_c(\ell), \tag{36}$$

and in general $U(x) = 1$. The details of the correction procedure are explained in Appendix B. Note that the form (35) is difficult to apply in the WR, because the CPDF naturally tends to the Gaussian limit as $\overline{\xi}_2$ reaches values much smaller than unity. Our method is thus practically valid only in the nonlinear regime, so when we correct for finite volume effects, we sample scales that satisfy $\ell > \ell_0$ (see column g of Table 1). At the early stages of the simulations, where the correlation length is still a very small fraction of the box size, the correction for finite volume effects does not yield any significant change in the measured values of $S_Q$ (as expected) and is thus not necessary. Table 1 lists the values of $a$ for which finite volume effects have been corrected for (column h).

Figure 4 gives the measured values of $S_Q$, but now corrected for finite volume effects if needed, in the "reliable" scaling range $\ell_m \leq \ell \leq \ell_M$ (the values of the scales $\ell_m$ and $\ell_M$ are listed in Table 1). The agreement with self-similarity is considerably better than in Figure 3. The two $n = -1$ simulations, that were giving quite different values of $S_Q$ for the uncorrected measurements at late stages are now in striking agreement.

### 4.2.3 Measurements versus theoretical predictions

From Figure 4, one can see a good, although not perfect, agreement of the measurements with the predictions of perturbation theory (long dashes) in the WR, except when $n = +1$, a case we further discuss below. Generally, the open squares on the extreme left, which correspond to the largest sampled scale in the first snapshot of each simulation, seem to be spurious. This is certainly due to finite volume effects (again) and to the fact that such weakly nonlinear scales have not yet relaxed; i.e. they are still influenced by transients, as discussed in § 3.2.

In the $n = +1$ case, the equations (29), (30) and (31) are certainly not valid, because nonlinear effects may never be negligible, even when $\overline{\xi}_2 \ll 1$ (this is true at least for $S_3$, see, e.g., JBC). Moreover, the effective realization of the initial power-spectrum is truncated to white noise at the Nyquist frequency (JBC) and no power exists on scales larger than the box size, so the effective values of $S_Q$ in the WR should be scale-dependent functions, not necessarily close to the theoretical predictions (29), (30) and (31). When $\overline{\xi}_2$ is smaller, but close to unity, the measurements are in agreement with the perturbation theory predictions for $n = 0$, as argued by JBC. For smaller values of $\overline{\xi}_2$ it is difficult to make any statement, since the values of $S_Q$ we measure are quite noisy, small or negative. Note, interestingly, that these results agree roughly with equations (29), (30) and (31) taken with $n = +1$, since in that case the predicted values of $S_Q$ are small: $S_3 \simeq 0.86$, $S_4 \simeq 0.56$ and $S_5 \simeq 0.18$.

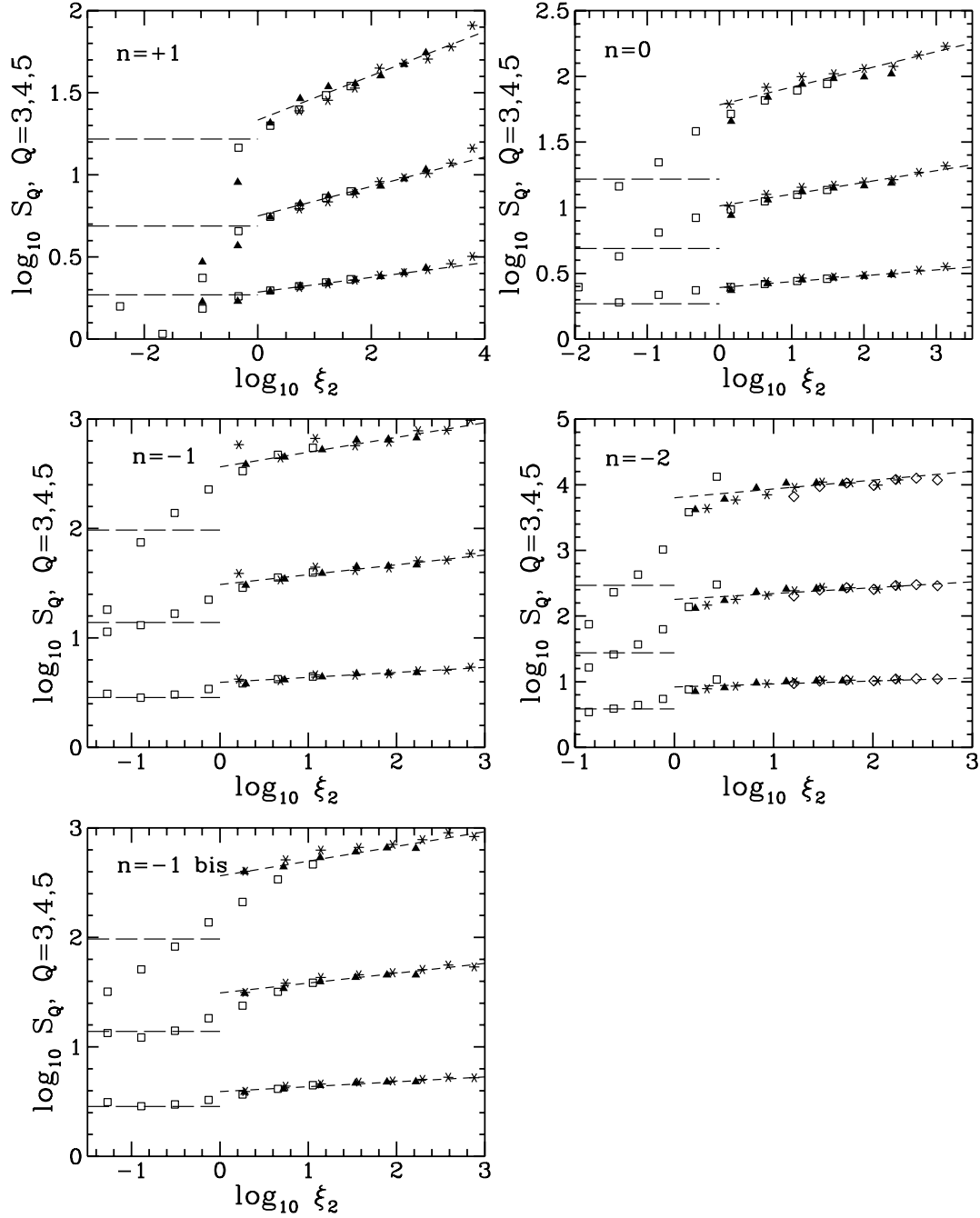

Figure 4: Same as Figure 3, but finite volume effects have been corrected for when necessary, by extending to infinity the large-$N$ exponential tail exhibited by the CPDF (cf. Fig. 1); only the reliable scales have been displayed, i.e. $\ell_\mathrm{m} \leq \ell \leq \ell_\mathrm{M}$, where $\ell_\mathrm{m}$ and $\ell_\mathrm{M}$ are listed in Table 1. Now, the agreement with self-similarity is much better: the curves corresponding to various expansion factors all superimpose, for a given value of $Q$ and $n$. The long dashes give the predictions (29), (30) and (31) from perturbation theory, valid in the limit $\overline{\xi}_2 \ll 1$. In the case $n = +1$, we display the predictions for $n = 0$ (see discussion in the text). The short dashes are the following phenomenological power-law fit $S_Q = \widetilde{S}_Q(\overline{\xi}_2/100)^{0.045(Q-2)}$, valid for $\overline{\xi}_2 > 1$, for all $n$. The values of $\widetilde{S}_Q$ are given in Table 3.

In the SR, the ratios $S_Q(\overline{\xi}_2)$ reach a plateau, although not exactly flat. Indeed, the functions $S_Q$ measured in each simulation in the nonlinear regime $\overline{\xi}_2 > 1$ are all compatible with the following power-law behavior

$$S_Q \simeq [D(\overline{\xi}_2)]^{Q-2} \widetilde{S}_Q, \quad Q = 3, 4, 5, \qquad (37)$$

with

$$D(\overline{\xi}_2) = \left(\frac{\overline{\xi}_2}{100}\right)^\delta, \qquad (38)$$

and

$$\delta \simeq 0.045. \qquad (39)$$

The values of $\widetilde{S}_Q$ are listed in Table 3. Note that they are quite close to those inferred from equations (29), (30) and (31) taken with some effective values $n_{\text{eff}}$ of the power-spectral index (see columns e and f of Table 3). This seems to indicate that the hierarchies of correlations in the SR are very similar to those given by perturbation theory, at least for small $Q$ (we neglect here the possible but small deviation from the scaling relation in equation [37]). The values of $\widetilde{S}_3$ are also in good agreement with the following phenomenological behavior $S_3 \simeq 9/(3 + n)$, as noticed by Fry, Melott & Shandarin (1993), who measured the three-body correlation function in Fourier space in simulations with $n$ ranging from $-3$ to $+1$.

The unexpected deviation from the scaling relation (5) reflected by equation (37) is very weak but impossible to deny, at least for $n = 0$ and $n = +1$. Indeed, let us denote by $S_Q^{\text{true}}$ the true values of $S_Q$ and by $S_Q^{\text{mes}}$ the measurements. The uncertainty in $S_Q^{\text{mes}}$ can be described as follows

$$S_Q^{\text{true}} = S_Q^{\text{mes}} f_Q^{\pm 1}, \qquad (40)$$

where $f_Q^{\pm 1}$ indicates that $S_Q^{\text{mes}}$ is to be multiplied by $f_Q$ or $1/f_Q$. The values of $f_Q$ are listed in Table 3. They were estimated by considering a range of possible and reasonable values of the parameters involved in equation (35) and are thus valid only in the nonlinear regime $\overline{\xi}_2 \gtrsim 1$. Note that the fact that we have two $n = -1$ realizations does not really help to make $f_Q$ closer to unity in that case: to have a significant improvement, we would need a few more simulations. However, we noticed that the measured exponential tails had quite similar features in each $n = -1$ simulation, so we used the same parameters for both samples in equation (35) (but the very details of the correction are slightly different, see Appendix B). So one should not over-interpret the very good agreement between the two $n = -1$ samples, once finite volume effects have been corrected for: part of this is due to our choice of the parameterization of the correction. However, if there are differences, they should be within $f_Q$, which corresponds to a small logarithmic shift, at most of order $\pm 0.08$, $\pm 0.13$, $\pm 0.2$ respectively for $S_3$, $S_4$ and $S_5$.

The dynamic range we have at our disposal in the nonlinear regime $\overline{\xi}_2 > 1$ is always at least two orders of magnitude for the function $\overline{\xi}_2$. Except for $n = -2$, this is still the case when one considers the regime $\overline{\xi}_2 \gtrsim 10$. However, for the more restrictive regime $\overline{\xi}_2 \gtrsim 100$, where $\overline{\xi}_2$ agrees with the power-law behavior given by equations (21) and (22), we have access to such a dynamic range only for $n = +1$.

Table 3: Measured values of $S_Q$

| $n^a$ | $Q^b$ | $\tilde{S}_Q{}^c$ | $f_Q{}^d$ | $n_{\text{eff}}{}^e$ | $S_Q^{\text{wea}}(n_{\text{eff}})^f$ | $S_Q^{\text{FMS}\,g}$ |
|---|---|---|---|---|---|---|
|      | 3 | 10.3  | 1.3  |      | 10.9  | 9.0  |
| -2   | 4 | 271   | 1.5  | -9   | 254   | –    |
|      | 5 | 11750 | 2.0  |      | 9294  | –    |
|      | 3 | 4.8   | 1.2  |      | 4.9   | 4.5  |
| -1   | 4 | 47    | 1.35 | -3   | 46    | –    |
|      | 5 | 679   | 1.55 |      | 656   | –    |
|      | 3 | 3.0   | 1.1  |      | 2.9   | 3.0  |
| 0    | 4 | 16    | 1.35 | -1   | 14    | –    |
|      | 5 | 113   | 1.65 |      | 96    | –    |
|      | 3 | 2.4   | 1.1  |      | 2.4   | 2.25 |
| +1   | 4 | 8.5   | 1.3  | -0.5 | 8.8   | –    |
|      | 5 | 40    | 1.65 |      | 45    | –    |

[a] spectral index.
[b] order $Q$ of the ratio $S_Q$.
[c] the measured value of $S_Q$ when $\overline{\xi}_2 = 100$ (see eq. [37]).
[d] the expected uncertainty factor in the measurements. If $S_Q^{\text{true}}$ is the true value of $S_Q$, and $S_Q^{\text{mes}}$ the value we measure, then $S_Q^{\text{true}} = S_Q^{\text{mes}} f_Q^{\pm 1}$.
[e,f] the "effective" spectral index and the expected values of $S_Q$ when one tries to use formulae (29), (30) and (31) to fit the measured $S_Q$ in the SR.
[g] The value of $S_3$ inferred from the measurements of Fry, Melott & Shandarin (1993).

The variations in $S_Q$ corresponding to a change of a factor 100 in $\overline{\xi}_2$ are, according to equation (37)

$$S_3(100\overline{\xi}_2)/S_3(\overline{\xi}_2) \simeq 1.2, \tag{41}$$
$$S_4(100\overline{\xi}_2)/S_4(\overline{\xi}_2) \simeq 1.5, \tag{42}$$
$$S_5(100\overline{\xi}_2)/S_5(\overline{\xi}_2) \simeq 1.85. \tag{43}$$

These ratios are close to unity but nevertheless larger than the estimated error factors $f_Q$ given in Table 3, except for $n = -2$. There is still an ambiguity for $n = -1$, but the cases $n = 0$ and $n = +1$ are undoubtedly in disagreement with the scaling relation. For $n = -2$ and marginally for $n = -1$, the measured $S_Q$ in the SR are compatible with the scaling relation.

To summarize the results of this section, we find a *weak but significant* deviation from the stable clustering hypothesis for the ratios $S_Q(\ell)$, $3 \leq Q \leq 5$ in the nonlinear regime, but our measurements are not yet conclusive for the cases $n = -1$ and $n - 2$.

# 5 Scaling behavior of the high-order correlations: the void probability

## 5.1 Self-similarity and spurious effects

### 5.1.1 Effects of self-similarity on $\hat{\sigma}$

We now want to see what the expected self-similar behavior implies for the void probability distribution function. To do so, let us rewrite equation (7) as

$$\hat{\sigma}(m, \ell, t) = \sum_{N \geq 1} (-1)^{N-1} \frac{S_N(\ell/s(t))}{N!} \left[ m\ell^3 \overline{\xi}_2(\ell/s(t)) \right]^{N-1}. \tag{44}$$

For a self-similar system, using the fact that $s(t) \propto a^\alpha$ (eqs. [14], [15]), this is equivalent to

$$\hat{\sigma}(m, \ell, a) = \hat{\sigma}(m\kappa^{-3}, \ell\kappa, a\kappa^{1/\alpha}), \tag{45}$$

where $\kappa$ is an arbitrary number and $a$ is the expansion factor. To check for self-similarity, we must thus measure the function $\sigma(m', \ell', a')$ at number densities $m' = m\kappa^{-3} \neq m$. Fortunately, it is possible to estimate the VPDF of a realization with an average number density $m' \neq m$ analytically once the function $P_N(m, \ell)$ is known. Indeed, we have (see, e.g., Hamilton 1985; Hamilton, Saslaw, & Thuan 1985; CBSII)

$$P_0(m, \ell) = \sum_{N=0}^{\infty} \left( 1 - \frac{m'}{m} \right)^N P_N(n, \ell). \tag{46}$$

Note that this series converges at least for $m'/m \leq 2$. It was shown by BH and CBSII that this formula gives quite accurate results in practice for $m'/m \lesssim 2$; it is difficult to use for $m'/m > 2$. In the case $m'/m \lesssim 1$, the VPDF given by equation (46) will be denoted as a "diluted" VPDF. It is equivalent to the VPDF measured in a subsample of average number density $m'$ randomly extracted from a parent sample of average number density $m$.

### 5.1.2 Grid effects

Figure 5 gives, in logarithmic coordinates, the quantity $\hat{\sigma}$ as a function of $\ell/s(t)$ measured in our simulations (except for one of the two $n = -1$ samples) for various expansion factors and various number densities. The symbols (already used in previous figures) represent the direct measurements in the $N$-body simulations, while the dashed, long-dashed (and dotted-dashed curves for $n = -2$) represent various diluted VPDFs using equation (46) (see the caption of Fig. 5). According to the self-similar solution, they should superimpose to the closest symbols, which is indeed the case, except perhaps for $n = -2$ and for the first snapshot of the $n = -1$ simulation. In these cases, the symbols tend to lie above the dashed curves. This is mainly due to grid effects.

Indeed, as discussed extensively in CBSII, the information carried by the regular pattern of particles used to set up initial conditions is likely to be conserved in underdense regions as the system evolves, especially if there is a cutoff at small scales in the power-spectrum (so grid effects should increase with $-n$). Consequently, the measured VPDF is smaller than it would be (so $\hat{\sigma}$ is larger) if the sample was locally Poissonian. The dashed curves are not very sensitive to such effects because they correspond to a diluted VPDF. Indeed the prescription given by equation (46) is equivalent to a random dilution if $m' \leq m$: if the dilution factor $\kappa^3$ is large enough, the information linked to the regular pattern will be destroyed. The factors $\kappa^3$ used in figure 5 are large, of order $(a_2/a_1)^{3\alpha} \sim 16$ or $(a_2/a_1)^{6\alpha} \sim 256$, where $a_2$ is the expansion factor corresponding to the parent VPDF and $a_1$ is the expansion factor corresponding to the diluted VPDF. According to CBSII, the VPDF directly measured in the $N$-body samples, which corresponds to the symbols in Figure 5 should be significantly influenced by grid effects when

$$P_0 < 1/e. \tag{47}$$

In Figure 5, the points that satisfy $P_0 < 1/e$ are circled. They indeed seem to be significantly further from the dashed curves for $n = -2$ and for the first snapshot of the $n = -1$ simulation (open squares). In other cases, grid effects appear to be unimportant. This is not surprising, because the $n = +1$ and $n = 0$ simulations have more "power" at small scales than the $n = -2$ and the $n = -1$ simulations: larger small scale power implies earlier shell crossing and phase mixing which destroy the information related to the grid (see also the discussion in § 3.2).

Note that some symbols at the extreme left of each panel also tend to lie above the dashed curves. We do not fully understand this small effect, but suspect it is linked to the short range softening of the forces. Indeed, we argued in § 4.1 that such an effect could influence $\overline{\xi}_2$ up to scales of order $\log_{10} \ell \sim -2.5$, making the system less clustered than it should be at small scales, i.e. closer to a Poisson distribution for which $\sigma$ is unity.

### 5.1.3 Finite sample effects

The vertical segments in each panel of Figure 5 are error-bars. According to CBSII, the error in $\hat{\sigma}$ can be estimated from

$$\frac{\Delta \hat{\sigma}}{\hat{\sigma}} \simeq \left| \frac{1}{m\ell^3 \hat{\sigma}} \frac{\Delta P_0}{P_0} - \frac{\Delta m}{m} \right|, \tag{48}$$

with

$$\left( \frac{\Delta m}{m} \right)^2 = \frac{1}{mL_{\rm box}^3} + \tilde{\xi}_2. \tag{49}$$

and

$$\left( \frac{\Delta P_0}{P_0} \right)^2 \simeq \frac{\overline{N}}{mL_{\rm box}^3} \frac{1 - P_0}{P_0} - 2\overline{N}^2 \frac{\partial \hat{\sigma}}{\partial N_{\rm c}} \tilde{\xi}_2. \tag{50}$$

The quantity $\partial \hat{\sigma}/\partial N_{\rm c}$ stands for the partial derivative of $\hat{\sigma}(N_{\rm c}, m)$ with respect to $N_{\rm c}$ (at fixed $m$) and $\tilde{\xi}_2$ stands for the average of $\xi_2$ within the sampled volume $V_{\rm box} \equiv L_{\rm box}^3$:

$$\tilde{\xi}_2 \equiv \frac{1}{V_{\rm box}^2} \int_{V_{\rm box}} \xi_2(|\mathbf{r}_1 - \mathbf{r}_2|) d^3 r_1 d^3 r_2. \tag{51}$$

We approximate this quantity as the integral of $\xi_2(|\mathbf{r}_1 - \mathbf{r}_2|)$ in the sphere of volume $V_{\rm box}$:

$$\tilde{\xi}_2 \simeq \overline{\xi}_2 \left( (3/4\pi)^{1/3} L_{\rm box} \right). \tag{52}$$

We refer to CBSII for the derivation and a discussion of the meaning and validity of equations (48), (49) and (50).

There is a transition scale $\ell_{\rm cut}$ above which $\Delta P_0/P_0$ (and $\Delta \hat{\sigma}/\hat{\sigma}$) becomes suddenly much larger than unity. This scale is simply defined by (CBSII)

$$P_0(m, \ell_{\rm cut}) \frac{L_{\rm box}^3}{\frac{4\pi}{3}\ell_{\rm cut}^3} \equiv 1. \tag{53}$$

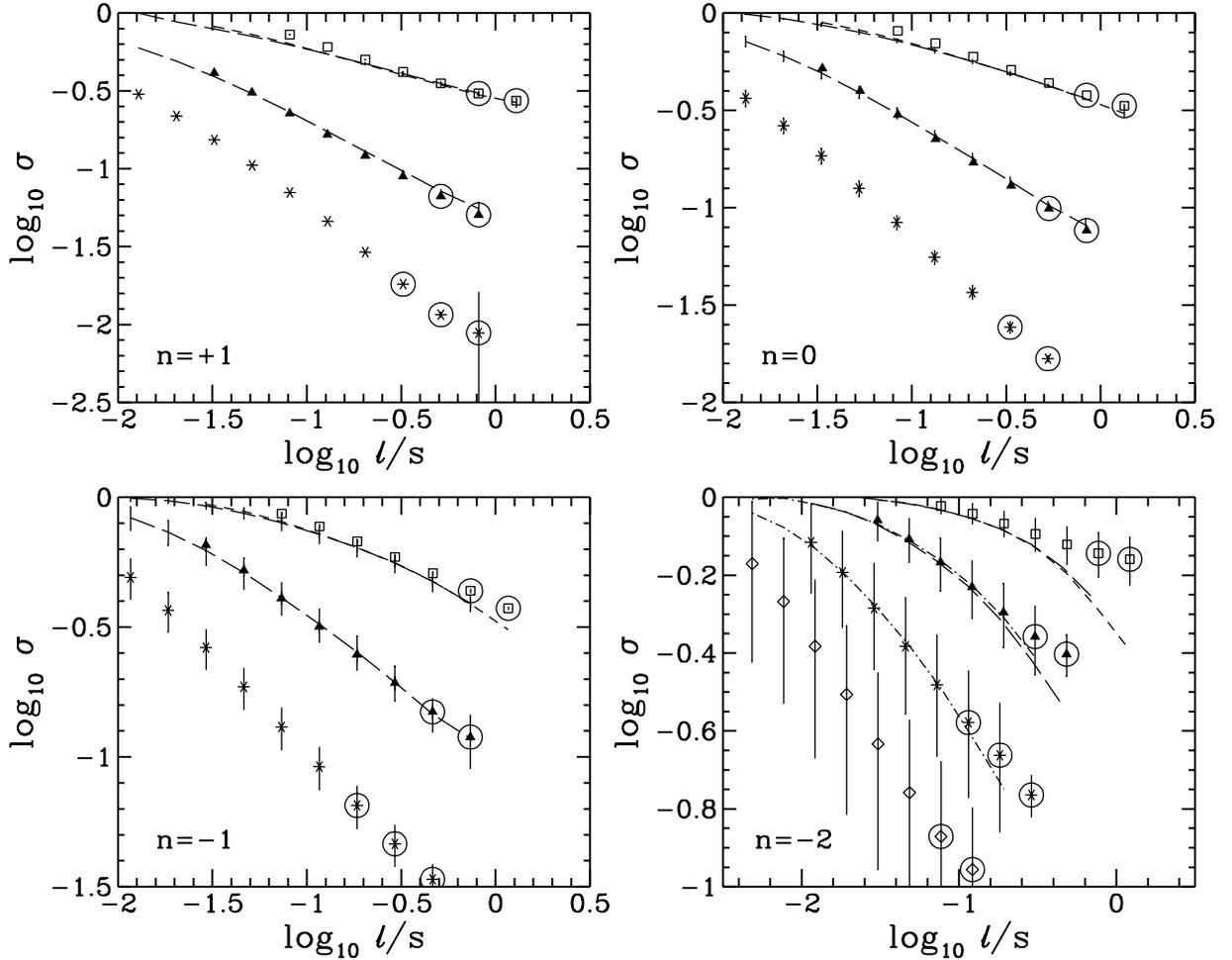

Figure 5: The quantity $\log_{10} \hat{\sigma}(m, \ell)$ as a function of $\log_{10} \ell/s$, measured in each simulation for various expansion factors (for symbols, see Table 1) in the available scale range (eq. [11]). In each panel, the short-dashed curve represents an analytic dilution by a factor $\kappa^3$ of the VPDF corresponding to the triangles using formula (46) so that, according to the self-similar solution (see eq. [45]), it should superimpose on the open squares. Similarly, the two long-dashed curves represent analytical dilutions of the VPDF corresponding to the stars. They should superimpose on the triangles and the open squares. The two dotted-dashed curves in the bottom right panel represent analytical dilutions of the VPDF corresponding to the diamonds. They should superimpose on the stars and the triangles. The circled points satisfy $P_0 < 1/e$. The vertical segments are error-bars estimated by using equation (48). For the sake of legibility, we put the error-bars for the case $n = -2$ only on the direct measurements, i.e. the symbols. The points such that $\Delta \hat{\sigma}/\hat{\sigma} \geq 1$ have been removed.

Above this scale, there is typically only one "statistically independent" empty cell: if $\ell \gtrsim \ell_{\rm cut}$ the VPDF is dominated by the largest void in the sample and its measurement is meaningless. We thus removed from Figure 5 the points satisfying $\ell \gtrsim \ell_{\rm cut}$.

To estimate the value of $\tilde{\xi}_2$ we use linear theory (eq. [20]). We can notice that the error-bars are insignificant for $n = 0$ and $n = +1$, and quite reasonable for $n = -1$. However, they are large in the $n = -2$ simulation, which is not surprising. Indeed, in that case (and in the case $n = -1$), the errors are, except for the largest scales, dominated by the terms in $\tilde{\xi}_2$ in equations (49) and (50). Now, the large scale "power" increases with $-n$, so do the values of $\tilde{\xi}_2$ (at fixed correlation length). For example, for $\ell_0 \simeq 0.13$ (see Table 1) one finds $\tilde{\xi}_2 \simeq 2 \times 10^{-3}$, $1.5 \times 10^{-2}$, $6 \times 10^{-2}$ and $0.18$ respectively for $n = +1$, $0$, $-1$ and $-2$.

Note, however, that we overestimate the finite sample error in our case from equations (48), (49) and (50). These equations assume that our $N$-body simulations are subsamples of much bigger sets with fluctuations on scales larger than $L_{\rm box}$. These fluctuations not only induce some uncertainty in the measured VPDF, but also in the average number density $m$ through equation (49). In our $N$-body samples, we know the average number density exactly, so the real error in our measurement is probably smaller than the one we compute by using the above prescription. For example, the difference between the values of $\hat{\sigma}(m, \ell, a)$ measured in our two $n = -1$ simulations are extremely small, much smaller than suggested by the error-bars of the bottom left panel of Figure 5 (however, for our example to be fully convincing, more $n = -1$ realizations would be needed). Also, the agreement with self-similarity in the $n = -2$ case is excellent, well within the error-bars, except for the first snapshot (squares) and the largest scales, but we know that this is mainly due to grid effects.

It is anyway not appropriate to set $\Delta m/m = 0$ instead of the value given by equation (49) and still use equation (50), because the VPDF is statistically correlated with $m$. By doing that, we would overestimate the errors even more.

### 5.2 Void probability and scaling relation

The above study of $\hat{\sigma}(n, \ell, t)$ enabled us to examine self-similarity in detail to determine a dynamic range over which the measurements can be trusted. Now, we check whether the scaling relation is verified. If this were the case, $\sigma(n, \ell, t)$ would scale as a function $\sigma(N_{\rm c})$, as discussed in the introduction. In this system of coordinates, however, it is not easy to distinguish the WR from the SR, so these two regimes should be studied separately. Contrary to the regime $\overline{\xi}_2 > 1$, the range of values of $\overline{\xi}_2$ we probe in the regime $\overline{\xi}_2 < 1$ is small, about one order of magnitude. Moreover, the measured VPDF is likely to vanish in this regime, corresponding to large cells that are likely to be always occupied. Simulations different from those described here would be needed to probe the WR. Therefore, we chose not to consider the WR in what follows.

#### 5.2.1 Void probability in the nonlinear regime

Figure 6 displays the quantity $\hat{\sigma}$ as a function of $N_{\rm c}$ measured in our simulations at various expansion factors $a$ and various number densities $m$. The curves plotted correspond to the same values of $(m, a)$ in Figure 5. We have selected the following scale range

$$\ell \geq \ell_{\rm m}, \quad \ell < \ell_{\rm cut}, \quad \ell < \ell_0, \tag{54}$$

where $\ell_{\rm m}$ is the "minimum reliable scale" defined in § 3 (see Table 1), and $\ell_{\rm cut}$ is defined by equation (53). For the direct measurements, represented by the symbols, we moreover imposed the constraint $P_0 \geq 1/e$ to avoid possible grid effects. This last condition is rather conservative. The results of § 5.1.2 indeed indicate that grid effects on the VPDF are negligible for the cases $n = 0$ and $n = +1$ and are rather small in the case $n = -1$, except for the smallest value of $a$.

The vertical error-bars are the same as those computed in § 5.1.3. Rigorously, we should also have horizontal error-bars that would account for the uncertainty on the measurements of $N_c = \overline{N}\,\overline{\xi}_2$. But the analysis of $\overline{\xi}_2$ in § 4.1 suggests that such uncertainties are rather small for $\ell \geq \ell_m$, except for $n = -2$. In this latter case, $\overline{\xi}_2$ is seriously affected by finite volume effects, and we propose a way to extract the physical information from the measurements that we use here: the number $N_c$ used to construct the bottom right panel of Figure 6 is not the measured one, but rather the quantity

$$N_c^{\text{fit}} = \overline{N}\; 10^{F[\log_{10}(\ell/s)]}, \tag{55}$$

where $F$ is the function defined by equations (27) and (28).

Note that for the case $n = -1$, for which we have two realizations, the direct measurement of the quantity $\hat{\sigma}$ as a function of $N_c$ gives results that are in excellent agreement for the two samples, except perhaps for the stars in the bottom left panel of Figure 6 that correspond to the largest expansion factor analyzed, $a = 16$. In that case, as discussed in § 4.1, $\overline{\xi}_2$ is affected by finite volume effects and one can detect a small difference between the simulations. This would correspond to a maximum horizontal shift on the bottom left panel of Figure 6 smaller than the size of the symbols used to make the plots so that the measurement in the other simulation is well inside the region defined by the vertical error-bars.

Clearly, for $n = +1$ and $n = 0$, $\hat{\sigma}(N_c, m, a)$ does not scale as a function of a single variable $\sigma(N_c)$. The case $n = -1$ is more ambiguous, but the deviation from scale-invariance is of the same order as the error-bars which we know overestimate the true errors. Although the measurements are certainly of much better quality than is suggested by the error-bars in Figure 6, the case $n = -2$ is overly influenced by finite volume effects to support a definitive conclusion: the measurements are compatible with scale invariance as well as with a slight deviation from it.

These results confirm those of § 4, which means that the behavior of the high order correlation hierarchy qualitatively follows the one of the low-order correlation hierarchy.

### 5.2.2 Fine tuning

How does the scaling behavior of the VPDF agree with the scaling behavior of the low order correlations? In other words, is it possible to generalize the power-law description (37) to any order $Q$? If this were the case, we would have

$$\hat{\sigma}(n, \ell) = 1 - \frac{1}{D(\overline{\xi}_2)} + \frac{1}{D(\overline{\xi}_2)} \widetilde{\sigma}(D(\overline{\xi}_2) N_c) \tag{56}$$

where

$$\widetilde{\sigma}(N_c{}') = \sum_{N \geq 1} (-1)^{N-1} \frac{\widetilde{S}_N}{N!} N_c{}'^{N-1}. \tag{57}$$

In other words, if equation (37) was valid for any $Q$, the quantity $\widetilde{\sigma} = D(\hat{\sigma} - 1) + 1$ would depend only on the number

$$N_c{}' \equiv D(\overline{\xi}_2) N_c. \tag{58}$$

This behavior is not what we find. This failure is not surprising, because equation (56) is quite likely to be unrealistic. Indeed, the condition $P_0 \leq 1$ implies $\hat{\sigma} \geq 0$. Moreover, using $\hat{\sigma}(x, y) =$ $\sum_{N\geq 1}(-1)^N(S_N(y)/N!)x^{N-1}$, we have $\widetilde{\sigma}(N'_c) = \hat{\sigma}(x \equiv N'_c, y = \ell_{100})$, where $\ell_{100}$ is the scale at which $\overline{\xi}_2 = 100$ or equivalently $D(\ell_{100}) \equiv 1$. So $\widetilde{\sigma}$ must be also positive, which implies, if $\hat{\sigma} \leq 1$, that

$$D \leq \frac{1}{1-\hat{\sigma}}, \qquad (59)$$

a condition which is not likely to be always fulfilled, particularly when $\ell > \ell_{100}$ (so $D > 1$) and $N_c$ is large. Indeed, in the regime $N_c \gg 1$, we find that $\hat{\sigma}$ is always much smaller than unity in the reliable dynamic range (so we should have $D \lesssim 1$ in this regime).

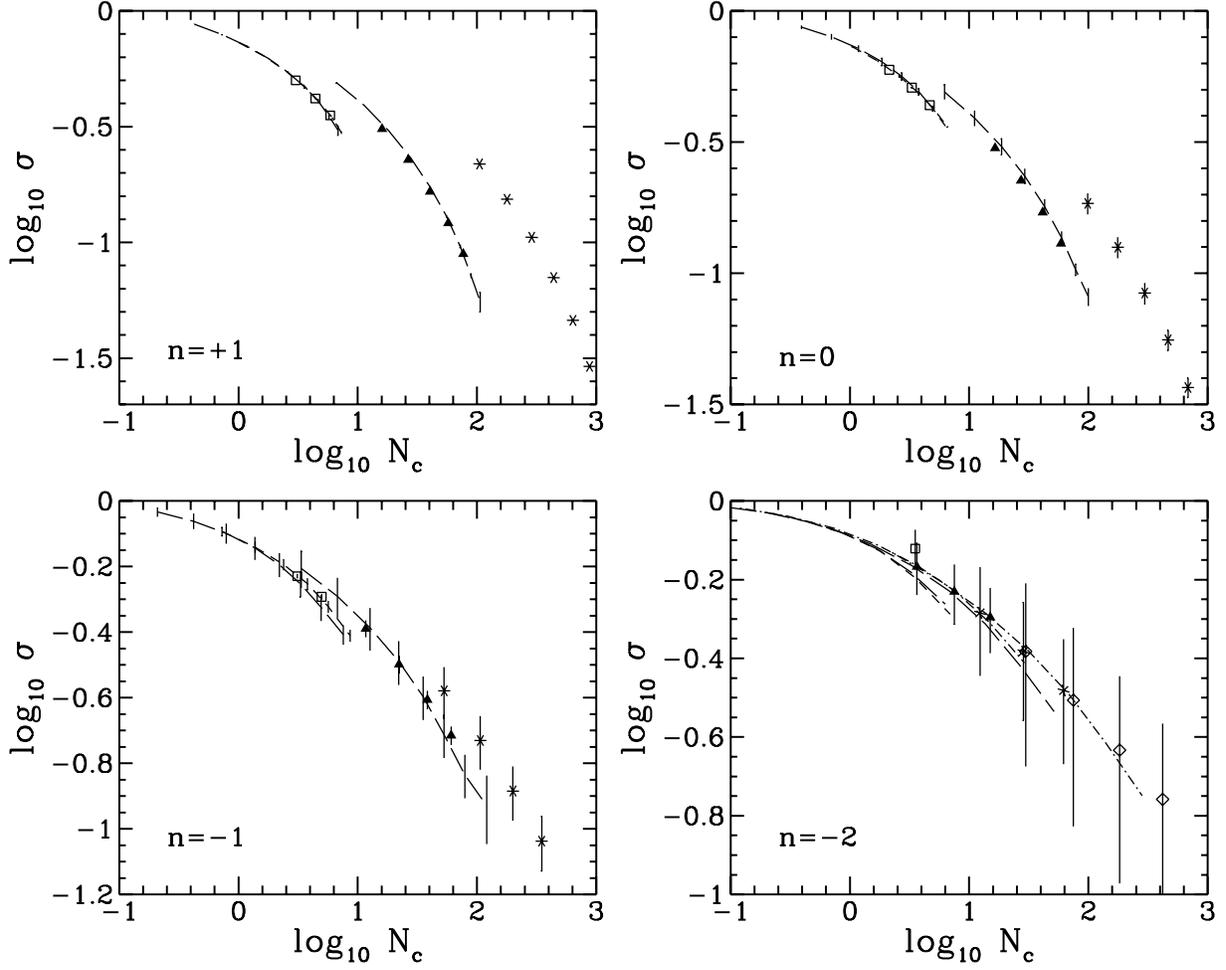

Figure 6: Same as Figure 5, but the quantity $\hat{\sigma}$ is represented as a function of $N_c$ in the scaling range (54). Only the nonlinear scales $\ell > \ell_0$ have been selected. For the direct measurements (symbols), the points satisfying $P_0 < 1/e$ have been removed (because of possible grid effects).

Thus, the tempting generalization of equation (37) to high-order correlations is not warranted.

## 6 Discussion, conclusions

### 6.1 Main results

We have measured the count probability distribution function (CPDF) in a set of five high resolution $N$-body simulations of self-gravitating expanding universes with scale-free initial power-spectra $\langle|\delta_k|^2\rangle \propto k^n$, $n = -2, -1$ (two realizations), 0 and $+1$. We focussed on the measurement of low-order averaged correlation functions $\overline{\xi}_Q$, $Q \leq 5$ as well as high-order correlations through the void probability distribution function (VPDF). Our primary goal was to determine whether or not the expected scaling relation $S_Q \equiv \overline{\xi}_Q / \overline{\xi}_2^{Q-2}$ = constant with respect to scale was achieved in the regime $\overline{\xi}_2 \gg 1$, as expected if the stable clustering hypothesis applies. In our analyses, we carefully studied spurious effects such as grid contamination, loss of dynamics due to the short range softening of the forces, and sampled volume finiteness. Our main conclusions are:

1. Finite volume effects are important for low-order correlations when $Q \geq 3$, increasing with $-n$. Fortunately one can correct for such errors, as suggested by CBSI. We performed such corrections, by extrapolating to infinity the exponential tail exhibited by the CPDF at large $N$. Finite volume effects were so large in the $n = -2$ case in our $N$-body data that they strongly affected the $Q = 2$ measurement ($\overline{\xi}_2$), which was not the case for $n \geq -1$. The VPDF was seen to be weakly sensitive to such defects, as expected (see CBSII).

   Grid effects, due to the fact that the simulations were started from a slightly perturbed regular pattern of particles, are small for the low-order correlations: they affect only the smallest scales at early stages of the simulations, typically $\ell \lesssim \ell_c$ where $\ell_c$ is the typical distance between two particles in overdense regions defined by $N_c(\ell_c) \equiv 1$. Grid effects on the VPDF (CBSII) are significant only when $n \leq -1$ and are of course stronger at early stages. They are quite strong in the case $n = -2$.

   In our simulations, the contamination due to the short range softening of the forces typically affected the measurement of the CPDF up to scales as large as $4\varepsilon$, where $\varepsilon$ is the softening length.

2. The quantities $S_Q$, $3 \leq Q \leq 5$ considered as functions of $\overline{\xi}_2$, exhibit two plateaus separated by a smooth transition around $\overline{\xi}_2 \sim 1$. The measured values of $S_Q$ in the weakly nonlinear regime $\overline{\xi}_2 \lesssim 1$ (WR) are in reasonable agreement with the perturbation theory predictions. The values of $S_Q$ measured in the strongly nonlinear regime SR, when $\overline{\xi}_2 > 1$, are larger than those measured in the WR. The difference actually increases with $-n$. The value of $S_Q$ at the scale where $\overline{\xi}_2 = 100$ can be well approximated by the values predicted by perturbation theory, but with an effective index $n_{\text{eff}} = -9, -3, -1, -0.5$ respectively for $n = -2, -1, 0, +1$. The strongly nonlinear statistics thus behaves in a manner similar to the WR. Note also that the measurement of $S_3(\overline{\xi}_2 = 100)$ is in rough qualitative agreement with the findings of Fry, Melott & Shandarin (1993), namely that $S_3 \sim 9/(3+n)$.

3. The plateau in the nonlinear regime $\overline{\xi}_2 > 1$ is however not exactly flat, but rather of the form

$$S_Q \propto \overline{\xi}_2^{\delta(Q-2)}, \quad \delta \simeq 0.045 \quad \text{for } 3 \leq Q \leq 5, \tag{60}$$

with $\delta \simeq 0.045$, independently of $n$. However, because of the remaining uncertainties in our measurements, the case $n = -2$ is also compatible with the scaling relation in the regime $\overline{\xi}_2 \gtrsim 10$. This is also true for $n = -1$, but only marginally.

4. The scaling behavior of the VPDF is in qualitative agreement with the above results: a significant deviation from the scaling relation is found in the nonlinear regime for the cases $n = 0$ and $n = +1$. With our error-bars, that overestimate the true errors, we see that our simulated $n = -1$ distribution agrees only marginally with the scaling relation. The measurements in the $n = -2$ simulation are too greatly affected by finite volume effects to enable firm conclusion. We found that equation (60) could not be generalized to arbitrary $Q$.

## 6.2 Discussion

The general trend seen in our measurements is in rough agreement with the similar work of Lucchin et al. (1994), but with differences in detail. Indeed, our estimates are more accurate, because we better control various spurious effects discussed in point 1 above, particularly finite volume effects. Moreover, because of our high spatial resolution, we have access to a larger effective range of values of $\overline{\xi}_2$, approximately one order of magnitude larger than these authors, who used a larger number of particles $N_{\text{par}} = 2\,192\,152$ than us, but evolved with a low resolution $N$-body code to perform their simulations. However, we hardly probe the weakly nonlinear regime $\overline{\xi}_2 \lesssim 1$. Even if the comparisons of our measurements with perturbation theory are satisfying, our $N$-body simulations are not really appropriate in that regime. More suitable investigations of the weakly nonlinear regime have been undertaken by Gaztañaga & Baugh (1995) for $(\Omega, \Lambda) = (1, 0)$ and $(0.2, 0.8)$ CDM simulations ($\Lambda$ is the cosmological constant) and they find excellent agreement between theory and measurements of $S_Q$ up to $Q = 10$.

The very weak deviation from the scaling relation we find in the strongly nonlinear regime is compatible with the earlier work of CBSI on matter distributions coming from CDM, white noise and HDM simulations. But it seems to disagree with some recent results in the literature, particularly those of Lahav et al. (1993, LIIS). If we were to believe the results of LIIS, we should have, in equation (60), $\delta \sim 0.2$ for low-density CDM and $\delta \sim 0.1$ for $n = 0$, which corresponds to much stronger deviations from scale-invariance than we measure. We think that the measurements of LIIS are contaminated by finite volume effects, particularly in the low-density CDM case. Our analyses, as stated by CBSI, indeed show that finite volume effects are more important when $-n$ increases, or when there is a cut-off at small scales in the initial density power-spectrum, which is the case for a CDM spectrum when compared, for example, to a white noise spectrum. Similarly, Suto & Matsubara (1994) measured the three-body and four-body correlation functions $\xi_3$ and $\xi_4$ (in the same simulations as LIIS) and computed various parameters associated with the hierarchical model (e.g., Groth & Peebles 1977; Fry 1984b) at these orders: the extrapolation of their results to our averaged correlation functions suggest $\delta \sim 0.1$ for low-density CDM and $\delta \sim 0.05$ for $n = 0$. Their measurements, although of much better quality than LIIS, are certainly still contaminated by finite volume effects in the CDM case, but are in agreement with our analyses for white noise initial conditions. More recently, Bonometto et al. (1995) have measured $S_3$ and $S_4$ in mixed dark matter and in CDM $N$-body samples and found strong deviation from scale-invariance. They attributed it to finite volume effects. Indeed, although their simulations involve a very large number of particles, more than $10^7$, their measurements are strongly contaminated by finite volume effects, mainly because the simulation box size they use ($L_{\text{box}} = 100$ Mpc) is not very large compared to the correlation length ($\ell_0 \gtrsim L_{\text{box}}/10$).

To conclude this paper, as clearly illustrated by Figures 4 and 6, we establish the existence of a *very weak* but *significant* deviation from the scaling relation (5) in the part of the strongly nonlinear regime (SR) we probe, at least for $n > -1$. It is however important to recognize that our measurements do not completely reject the stable clustering hypothesis, which argues that the scaling relation should hold in the regime $\overline{\xi}_2 \gg 1$ if a self-similar solution is reached. Indeed, earlier measurements by EFWD show that the power-law behavior (21) of $\overline{\xi}_2$ implied by the stable clustering hypothesis is achieved only when $\overline{\xi}_2 \gtrsim 100$, as illustrated by Figure 2. In this regime, we hardly probe an order of magnitude of values of $\overline{\xi}_2$ for $n = -1$ and $n = -2$, a bit more for $n = 0$ and two orders of magnitude for $n = +1$. With this more restrictive approach, the deviation from the scaling relation we measure is really significant only in the case $n = +1$, and it is anyway quite small. One would need much larger high resolution simulations to significantly probe the strongly nonlinear regime $\overline{\xi}_2$, with large number of particles (say more than $128^3$ instead of our $64^3$) and higher spatial resolution (smaller short range softening length $\varepsilon$).

**Acknowledgments.** This work was supported in part by the Pittsburgh Supercomputing Center and the National Center for Supercomputing Applications, in part by an allocation from the scientific counsel of IDRIS, Palaiseau, and in part by the National Science Foundation under Grant No. PHY94-07194. SC is supported by DOE and by NASA through grant NAG-5-2788 at Fermilab. Part of this work was done while SC was at the Institut d'Astrophysique de Paris (CNRS), supported by Ecole Polytechnique. LH is supported by the Alfred P. Sloan Foundation, NASA Theory Grant NAGW–2422, NSF Grants AST 90–18526 and ASC 93–18185, and the Presidential Faculty Fellows Program. This work was completed by the three authors at the Institute for Theoretical Physics (University of California, Santa Barbara).

# APPENDIX

## A  Timesteps

Following Efstathiou et al. (1985, EDFW), we use the following time variable:

$$p = a^\alpha, \quad \alpha = \frac{2}{n+3}, \tag{61}$$

where $a$ is the expansion factor, taken to be unity at the beginning of the simulation. In the units of the treecode of Hernquist et al. (1991), the gravitational constant is $G = 1$ and the mass of a particle is $m = N_{\text{par}}^{-1}$. To probe the same dynamic range as EFWD, all the simulations were evolved from $p = 1$ to $p_{\max} = 16$, except for $n = -2$ where $p_{\max} = 64$. The corresponding values of the expansion factor are given in Table 1.

According to EDFW, the timestep $\Delta p$ should satisfy the following requirement

$$\Delta p \lesssim \frac{1}{2}\alpha \dot{a} a^{\alpha+1/2} \left[\frac{\eta^3}{Gm}\right]^{1/2}, \qquad (62)$$

at any time for the time-centered leapfrog scheme to be stable against round-off errors, where $\eta$ is the softening parameter used by EDFW in their P$^3$M code. This $\eta$ corresponds to an effective softening length (EFWD)

$$\varepsilon \simeq \eta/3 \qquad (63)$$

for a potential of the form

$$\phi \propto (r^2 + \varepsilon^2)^{-1/2}. \qquad (64)$$

Although, we use a slightly different softening of the short range forces (i.e. a cubic spline interpolation; see Hernquist & Katz 1989; Goodman & Hernquist 1991), we assume that eq. (63) is still valid in our case. Equation (62) can now be rewritten

$$\Delta p \lesssim 1.73 \alpha a^\alpha (\varepsilon/\lambda_{\text{par}})^{3/2}, \qquad (65)$$

where $\lambda_{\text{par}}$ is the mean interparticle distance.

For $\varepsilon = \lambda_{\text{par}}/20$, which is the value we choose for the softening parameter, we find

$$\Delta p \lesssim 0.0194 \alpha a^\alpha, \qquad (66)$$

which leads to $\Delta p \lesssim 0.0097, 0.013, 0.019, 0.039$ for $n = 1, 0, -1, -2$ respectively. Such values also apply to the scale invariant simulations of EFWD and are indeed in very good agreement with the values of $\Delta p$ displayed on Table 1 of EFWD. Our choice of timestep was $\Delta p = 0.0084, 0.011, 0.017, 0.034$ respectively for $n = 1, 0, -1, -2$, which is slightly smaller than the one imposed by the condition (66).

# B  Correcting for finite volume effects

This paragraph explains in detail how we use the phenomenological form (35) to correct for finite volume effects.

Since $\bar{\xi}_2$ is not affected by finite volume effects, except for the case $n = -2$ which requires a special treatment as explained below, we use the measured values of $N_c$ in equation (35). We replace the measured CPDF by the analytical form (35) for (arbitrary) $N$ larger than a lower bound $N_{\text{crit}}(\ell)$. This number $N_{\text{crit}}$ must be large enough so that the function $P_N$ indeed exhibits the behavior implied by equation (35). Typically, we should have $N_{\text{crit}} \gtrsim$ a few $N_c$ when $U(x) = 1$ (but practically, we take $N_{\text{crit}} \gtrsim N_c$) and $N_{\text{crit}} \geq N_{\text{crit}}^{\min}$ with $N_{\text{crit}}^{\min} \gg 1$ (e.g., CBSI). The last condition is imposed because of discreteness effects. In some cases for $n = +1$ and $n = -2$, we used a more elaborate fit than equation (35) with $U = 1$, to have a description as accurate as possible of the shape of the CPDF. In that case, $N_{\text{crit}}$ can be smaller than $N_c$. Following BSD and BH, we take

$$U(x) = (1 + bx)^{-c}, \qquad (67)$$

Table 4: The values of the parameters used to correct for finite volume effects

| $n^a$ | $a^b$ | $|y_s|$ | $b$ | $c$ | $N_{\mathrm{crit}}^{\min}$ | $\log_{10}\ell$ | -2.6 | -2.4 | -2.2 | -2.0 | -1.8 | -1.6 | -1.4 | -1.2 | -1.0 |
|---|---|---|---|---|---|---|---|---|---|---|---|---|---|---|---|
| -2 | 3.2 | 18 | 0 | 1 | 40 | $N_{\mathrm{crit}}/N_c$ | – | 1.5 | 1.5 | 1.5 | 1.5 | 2.0 | 2.5 | – | – |
| | | | | | | $\eta$ | – | -2.6 | -2.6 | -2.6 | -2.6 | -2.8 | -3.0 | – | – |
| -2 | 5.2 | 18 | 0 | 1 | 100 | $N_{\mathrm{crit}}/N_c$ | – | 1.0 | 1.0 | 1.0 | 1.0 | 1.0 | 1.0 | 1.0 | 1.0 |
| | | | | | | $\eta$ | – | -2.45 | -2.45 | -2.45 | -2.45 | -2.45 | -2.45 | -2.57 | -2.68 |
| -2 | 8.0 | 18 | 3.7 | 0.75 | 200 | $N_{\mathrm{crit}}/N_c$ | – | 0.05 | 0.05 | 0.05 | 0.05 | 0.05 | 0.05 | 0.05 | – |
| | | | | | | $\eta$ | – | -1.7 | -1.7 | -1.7 | -1.7 | -1.7 | -1.7 | -1.7 | – |
| $-1^c$ | 6.4 | 5 | 0 | 1 | 40 | $N_{\mathrm{crit}}/N_c$ | – | 2.0 | 2.0 | 2.0 | 2.0 | 2.0 | 2.2 | – | – |
| | | | | | | $\eta$ | – | -2.1 | -2.2 | -2.1 | -2.2 | -2.35 | -2.7 | – | – |
| -1 | 16 | 5 | 0 | 1 | 40 | $N_{\mathrm{crit}}/N_c$ | – | 1.5 | 1.5 | 1.5 | 1.5 | 1.5 | 1.5 | 1.5 | 1.5 |
| | | | | | | $\eta$ | – | -2.1 | -2.1 | -2.1 | -2.1 | -2.1 | -2.2 | -2.35 | -2.7 |
| $-1^d$ | 16 | 5 | 0 | 1 | 40 | $N_{\mathrm{crit}}/N_c$ | – | 2.0 | 1.5 | 1.3 | 1.1 | 0.98 | 0.7 | 2.0 | 1.9 |
| | | | | | | $\eta$ | – | -2.1 | -2.1 | -2.1 | -2.1 | -2.1 | -2.2 | -2.35 | -2.7 |
| 0 | 16 | $\sim 2^e$ | 0 | 1 | 100 | $N_{\mathrm{crit}}/N_c$ | – | 2.5 | 2.5 | 2.5 | 2.5 | 2.3 | 3.0 | – | – |
| | | | | | | $\eta$ | – | -1.575 | -1.575 | -1.575 | -1.575 | -1.575 | -1.575 | – | – |
| | | | | | | $|y_s|^e$ | – | 2.0 | 2.0 | 2.0 | 2.0 | 1.8 | 1.5 | – | – |
| 0 | 64 | $\sim 2^e$ | 0 | 1 | 100 | $N_{\mathrm{crit}}/N_c$ | – | 3.0 | 2.0 | 2.0 | 2.0 | 1.0 | 1.0 | 1.0 | 2.0 |
| | | | | | | $\eta$ | – | -1.575 | -1.575 | -1.575 | -1.575 | -1.575 | -1.575 | -1.575 | -1.575 |
| | | | | | | $|y_s|^e$ | – | 2.2 | 2.0 | 2.0 | 2.0 | 2.0 | 2.0 | 1.8 | 1.5 |
| +1 | 256 | 1 | 0.486 | -1.35 | 800 | $N_{\mathrm{crit}}/N_c$ | 1.7 | 1.7 | 1.7 | 1.7 | 1.7 | 1.7 | 1.7 | 1.7 | 1.7 |
| | | | | | | $\eta$ | -1.35 | -1.35 | -1.35 | -1.35 | -1.35 | -1.35 | -1.35 | -1.35 | -1.35 |

[a] spectral index.
[b] expansion factor.
[c] the correction parameters are the same for both $n=-1$ simulations at $a=6.4$.
[d] the value of $N_{\mathrm{crit}}/N_c$ is slightly different for each of the $n=-1$ runs at $a=16$.
[e] in the case $n=0$, it appears appropriate to have scale dependent $|y_s|$ and constant $\eta$.

where $b$ and $c$ are *a priori* adjustable parameters that may vary with scale, but that can be fixed at constant values at the level of precision adopted here. To follow CBSI, since $\overline{\xi}_2$ is not affected by finite volume effects, except for $n=-2$, we impose a choice of the parameters such that the correction does not change the value of $\overline{\xi}_2$.

The case $n=-2$ is slightly more complicated. For $a=3.2$, where finite volume effects are still small for the function $\overline{\xi}_2$, we use the above procedure. For larger values of $a$, we use the fitting function $F(x)$ determined in § 4.1.3 to describe $\overline{\xi}_2$ and then write $N_c = \overline{N}\,\overline{\xi}_2^{\mathrm{fit}}$ in equation (35) with

$$\overline{\xi}_2^{\mathrm{fit}} = 10^{F[\log_{10}(\ell/s)]}. \tag{68}$$

We require that the measured function $\overline{\xi}_2$, once corrected for finite volume effects, be as close as possible to $\overline{\xi}_2^{\mathrm{fit}}$.

Table 4 lists the values of $N_{\mathrm{crit}}/N_c$, $N_{\mathrm{crit}}^{\min}$, $\eta(\ell)$, $|y_s(\ell)|$, $b$, $c$ we chose when we corrected for finite volume effects. Note that when $b=0$ and $c=1$, this simply means we took $U(x) \equiv 1$.

# References

Balian, R., & Schaeffer, R. 1989a, A&A, 220, 1 (BS)
Balian, R., & Schaeffer, R. 1989b, A&A, 226, 373
Baugh, C.M., Gaztañaga, E., & Efstathiou, G. 1994, MNRAS, 274, 1049
Bernardeau, F. 1992, ApJ, 392, 1
Bernardeau, F. 1994a, ApJ, 433, 1
Bernardeau, F. 1994b, A&A, 291, 697 (B94)
Bonometto, S.A., Borgani, S., Ghigna, S., Klypin, A., & Primack, J.R. 1995, MNRAS, 273, 101

Bouchet, F.R., & Lachièze-Rey, M. 1986, ApJ, 302, L37
Bouchet, F.R., Schaeffer, R., & Davis, M. 1991, ApJ, 383, 19 (BSD)
Bouchet, F.R., & Hernquist, L. 1992, ApJ, 400, 25 (BH)
Bouchet, F.R., Juszkiewicz, R., Colombi, S., & Pellat R. 1992, ApJ, 394, L5
Bouchet, F.R., Strauss, M.A., Davis, M., Fisher, K.B., Yahil, A., & Huchra, J.P. 1993, ApJ, 417, 36
Bouchet, F.R., Colombi, S., Hivon, E., & Juszkiewicz, R. 1994, A&A, 296, 575
Colombi, S., Bouchet, F.R., & Schaeffer, R. 1994, A&A, 281, 301 (CBSI)
Colombi, S., Bouchet, F.R., & Schaeffer, R. 1995, ApJS, 96, 401 (CBSII)
Davis. M., & Peebles, P.J.E. 1977, ApJS, 34, 425
Davis, M., & Peebles, P.J.E. 1983, ApJ, 267, 465
Efstathiou, G., Davis, M., Frenk, C.S., & White, S.D.M 1985, ApJS, 57, 241 (EDFW)
Efstathiou, G., Frenk, C.S., White, S.D.M., & Davis, M. 1988, MNRAS, 235, 715 (EFWD)
Fry, J.N. 1984a, ApJ, 279, 499
Fry, J.N. 1984b, ApJ, 277, L5
Fry, J.N., & Peebles, P.J.E. 1978, ApJ, 221, 19
Fry, J.N., Giovanelli, R., Haynes, M.P., Melott, A.L., & Scherrer, R.J. 1989, ApJ, 340, 11
Fry, J.N., Melott, A.L., & Shandarin, S.F 1993, ApJ, 412, 504 (FMS)
Gaztañaga, E. 1994, MNRAS, 286, 913
Gaztañaga, E., & Baugh, C.M. 1995, MNRAS, 273, L1
Goodman, J. & Hernquist, L. 1991, ApJ, 378, 637
Goroff, M.H., Grinstein, B., Rey, S.-J., & Wise, M.B. 1986, ApJ, 311, 6
Groth, E.J., & Peebles, P.J.E. 1977, ApJ, 217, 385
Hamilton, A.J.S. 1985, ApJ, 292, L35
Hamilton, A.J.S. 1988, ApJ, 332, 67
Hamilton, A.J.S., Saslaw, W.C., & Thuan, T.X. 1985, ApJ, 297, 37
Hernquist, L. & Katz, N.S. 1989, ApJS, 70, 419
Hernquist, L., Bouchet, F.R., & Suto, Y. 1991, ApJS, 75, 231
Hivon, E., Bouchet, F.R., Colombi, S., Juszkiewicz, R. 1994 A&A, in press (astro-ph/9407049)
Inagaki, S. 1976, Publ. Astron. Soc. Japan., 28, 77
Juszkiewicz, R., & Bouchet, F.R. 1992, Proceedings of the Second DAEC meeting, Meudon, France, Ed. G. Mamon & D. Gerbal
Juszkiewicz, R., Bouchet, F.R., & Colombi, S. 1993, ApJ, 412, L9 (JBC)
Lokas, E.L., Juszkiewicz, R., Weinberg, D.H., & Bouchet, F.R. 1994, preprint IASSNS-AST (astro-ph/9407095)
Lahav, O., Itoh, M., Inagaki, S., & Suto, Y. 1993, ApJ, 402, 387 (LIIS)
Lucchin, F., Matarese, S., Melott, A.L., & Moscardini, L 1994, ApJ, 422, 430
Matsubara, T., & Suto, Y. 1994, ApJ, 420, 497
Maurogordato, & S., Lachièze-Rey, M. 1987, ApJ, 320, 13
Maurogordato, S., Schaeffer, R., & da Costa, L.N. 1992, ApJ, 390, 17
Peebles, P.J.E. 1980, The Large Scale Structure of the Universe (Princeton University Press, Princeton, N.Y., U.S.A.) (LSS)
Schaeffer, R. 1984, A&A, 134, L15
Sharp, N.A. 1981, MNRAS, 191, 857
Sharp, N.A., Bonometto, S.A., & Lucchin, F. 1984, A&A, 130, 7
Suto, Y., & Matsubara, T. 1994, ApJ, 420, 504

Szapudi, I, Szalay, A., & Boschán, P. 1992, ApJ, 390, 350
Szapudi, I., & Szalay, A. 1993, ApJ, 408, 43
White, S.D.M. 1979, MNRAS, 186, 145
Zel'dovich, Ya. B. 1970, A&A, 5, 84